\documentclass[11pt]{article}              

\usepackage[margin=25mm]{geometry}
\usepackage{graphicx}
\usepackage{colortbl}
\usepackage{subfigure}
 
\usepackage{fancyhdr}
\usepackage{amssymb,amsfonts,amsmath,amsthm}
\usepackage{natbib}
\usepackage{bstnotations}

\usepackage[english]{babel} 
\usepackage{graphicx}
\usepackage{float}
\usepackage{color}
\usepackage{soul}
\usepackage{geometry}
\usepackage{natbib}
\usepackage{setspace}
\usepackage{caption}
\usepackage{multirow}

\usepackage{authblk}
	
\graphicspath{{./},{./Figures/}}

\newcommand{\cD}{\mathcal{D}}
\newcommand{\uF}{\overline{F}}
\newcommand{\lF}{\underline{F}}
\newcommand{\vt}{\ve{\theta}}
\newcommand{\vx}{\ve{x}}
\newcommand{\vX}{\ve{X}}
\newcommand{\vC}{\ve{C}}
\newcommand{\vc}{\ve{c}}
\newcommand{\lcm}{\underline{\cm}}
\newcommand{\ucm}{\overline{\cm}}
\newcommand{\cA}{\mathcal{A}}
\newcommand{\cY}{\mathcal{Y}}
\newcommand{\vCC}{\ve{\mathfrak{C}}}
\newcommand{\vcc}{\ve{\mathfrak{c}}}
\newcommand{\va}{\ve{a}}
\newcommand{\vT}{\ve{\Theta}}
\newcommand{\vZ}{\ve{Z}}

\newcommand{\vk}{\ve{k}}
\newcommand{\cK}{\mathcal{K}}

\begin{document}
\title{Uncertainty propagation of p-boxes using sparse polynomial chaos expansions}

\author[1]{R. Sch\"obi} \author[1]{B. Sudret} 

\affil[1]{Chair of Risk, Safety and Uncertainty Quantification,
  
  ETH Zurich, Stefano-Franscini-Platz 5, 8093 Zurich, Switzerland}

\date{}
\maketitle

\abstract{In modern engineering, physical processes are modelled and analysed using advanced computer simulations, such as finite element models. Furthermore, concepts of reliability analysis and robust design are becoming popular, hence, making efficient quantification and propagation of uncertainties an important aspect. In this context, a typical workflow includes the characterization of the uncertainty in the input variables. In this paper, input variables are modelled by probability-boxes (p-boxes), accounting for both aleatory and epistemic uncertainty. The propagation of p-boxes leads to p-boxes of the output of the computational model. A two-level meta-modelling approach is proposed using non-intrusive sparse polynomial chaos expansions to surrogate the exact computational model and, hence, to facilitate the uncertainty quantification analysis. The capabilities of the proposed approach are illustrated through applications using a benchmark analytical function and two realistic engineering problem settings. They show that the proposed two-level approach allows for an accurate estimation of the statistics of the response quantity of interest using a small number of evaluations of the exact computational model. This is crucial in cases where the computational costs are dominated by the runs of high-fidelity computational models.
\\[1em] 

 }

\maketitle

\section{Introduction} \label{sec:intro}

In modern engineering, computational simulations (\eg finite element-based simulations) have become a popular tool for predicting and analysing the behaviour of mechanical systems or engineering structures. The increasing knowledge in science and engineering leads to models of larger complexity, which requires increasing computational resources. At the same time, awareness on quantitative reliability, robustness, and design optimization is growing. Engineers are more and more concerned with the quantification of uncertainties \citep{SudretHDR,Derocquigny2012,Forrester2008}. 

In this context, a typical workflow consists of defining a computational model, determining a model for the uncertain input parameters, then propagating and analysing the uncertainty in the quantities of interest (QoI). The uncertainty in input parameters is traditionally quantified by probability theory, which describes the variability of a parameter by a single measure. However, in many situations probability theory is not appropriate to quantify completely uncertainty in the parameters. Indeed, aleatory uncertainty corresponds to the natural variability of an input parameter (non-reducible), whereas epistemic uncertainty is related to lack of knowledge which, in principle, could be reduced by gathering more information (\eg data points, measurements) \citep{Kiureghian2009}. 

A number of methods have been proposed to capture the characteristics of so-called \emph{imprecise probabilities} which include aleatory as well as epistemic uncertainties. Amongst those methods are Dempster-Shafer structures \citep{Dempster1967,Shafer1976}, possibility theory \citep{Dubois1988}, Bayesian hierarchical models \citep{Gelman2009}, fuzzy sets \citep{Moller2004}, probability-boxes (p-boxes) \citep{Ferson1996}, clouds \citep{Neumaier2004,Destercke2008}, and random sets \citep{Matheron1975,Molchanov2005}. The focus of this paper lies on p-boxes, which are defined by lower and upper boundary curves to the cumulative distribution function of an input variable.

In the context of probabilistic input, uncertainty propagation methods have been widely studied in the last decades through Monte Carlo simulation. However, when considering p-boxes in the input space, uncertainty propagation is more complex. A much lower number of methods have been developed for propagating p-boxes, amongst which are 
nested Monte Carlo algorithms \citep{Eldred2009,He2015} and 
interval-analysis-based algorithms \citep{Helton2004,Helton2004a}. These algorithms require a large number of model evaluations to ensure an accurate estimate of the uncertainty in the QoIs. Then, in the general case of expensive-to-evaluate models, these types of algorithms may become intractable. 

A popular strategy to reduce computational costs in uncertainty propagation is the use of meta-models in order to surrogate the exact model by an approximative, inexpensive-to-evaluate function \citep{SudretBookPhoon2015}. Common meta-modelling techniques are Gaussian process models (a.k.a Kriging) \citep{Santner2003,Krige1951}, Polynomial Chaos Expansions (PCE) \citep{GhanemBook2003} and support vector machines \citep{Gunn1998}. Traditionally, meta-modelling techniques have been used in the context of probabilistic input in a variety of problems such as uncertainty propagation \citep{BlatmanPEM2010,SchobiSudretIJ4UQ2015}, sensitivity analysis \citep{Sudret2008c}, structural reliability analysis \citep{Echard2011,Balesdent2013,SchobiASCE2015} and design optimization \citep{Dubourg2011a,MalikiSMO2016}. 

In contrast, meta-modelling techniques have only been used in few occasions to propagate uncertainties modelled by imprecise probabilities. Recent contributions include \cite{Chen2015b}, where fuzzy sets are propagated using generalized polynomial chaos meta-models, \cite{Hu2015}, where Kriging is used to estimate failure probabilities, and \cite{Li2012}, where epistemic uncertainty is propagated using polynomial surrogate models.

In this paper, we introduce a novel approach to propagate uncertainties modelled by p-boxes through complex computational models using sparse non-intrusive polynomial chaos expansions. 
The paper is structured as follows: Section~\ref{sec:impre} introduces the definition of p-boxes followed by the definition of two case studies inspired by engineering practice. Uncertainty propagation and related algorithms are discussed in Section~\ref{sec:free}. Section~\ref{sec:2meta} introduces the proposed approach, which makes use of meta-models at two levels of the uncertainty propagation workflow. Three applications (Section~\ref{sec:appl}) are used to illustrate and discuss the proposed method. The paper terminates with conclusions in Section~\ref{sec:conc}. 

\section{Imprecise probability} \label{sec:impre}
\subsection{Probability-boxes} \label{sec:pbox}

{Consider the probability space $(\Omega,\mathcal{F},\mathbb{P})$, where $\Omega$ denotes the outcome space equipped with the $\sigma$-algebra $\mathcal{F}$ and a probability measure $\mathbb{P}$. Let us denote by $X$ a random variable defined by the mapping $X:\, \omega\in\Omega \mapsto X(\omega)\in\cD_X\subset \Rr$, where $\omega\in\Omega$ is an elementary event and $\mathcal{D}_X$ is the support domain of $X$. A random variable $X$ is typically characterized by its \emph{cumulative distribution function} (CDF) $F_X(x) \eqdef \Prob{X\leq x}$ or, in case of continuous variables, by its \emph{probability distribution function} (PDF) $f_X(x) = \text{d}F_{X}(x) / \text{d}x$. }

{As seen in the definitions above, probability theory provides a single measure to quantify uncertainty in $X$. In other words, it is assumed that the uncertainty is known and quantifiable by a probability distribution through its CDF (and related PDF). In many cases, however, knowledge on $X$ is incomplete and probability theory is not sufficient to describe the uncertainty. This motivates the introduction of so-called \emph{probability-boxes} (p-boxes) which account for aleatory (natural variability) as well as for epistemic uncertainty in the description of variable $X$.}
 
Mathematically speaking, a \emph{p-box} is defined by lower and upper bounds to the CDF of $X$, denoted by $\lF_X$ and $\uF_X$ respectively \citep{Ferson1996,Ferson2004}. For any value $x\in \mathcal{D}_X$ the true but unknown CDF lies within these bounds, \ie $\lF_X(x)\leq F_X(x) \leq \uF_X(x), \ \forall x\in \mathcal{D}_X$. The boundary curves of the p-box mark extremes of the CDF and are thus themselves CDFs by definition. Note that this type of p-boxes is called \emph{free} p-boxes in the literature as opposed to parametric p-boxes, which are defined as a distribution the parameters of which are modelled by intervals \cite{Ferson2003,SchoebiICASP2015}. In this paper, only free p-boxes are discussed, because they are a generalization of parametric p-boxes. Hence in the remainder of the paper, the adjective \emph{free} will be omitted in the context of free p-boxes for the sake of simplicity. 

\begin{figure}[ht!]
\centering
\includegraphics[width=0.4\linewidth]{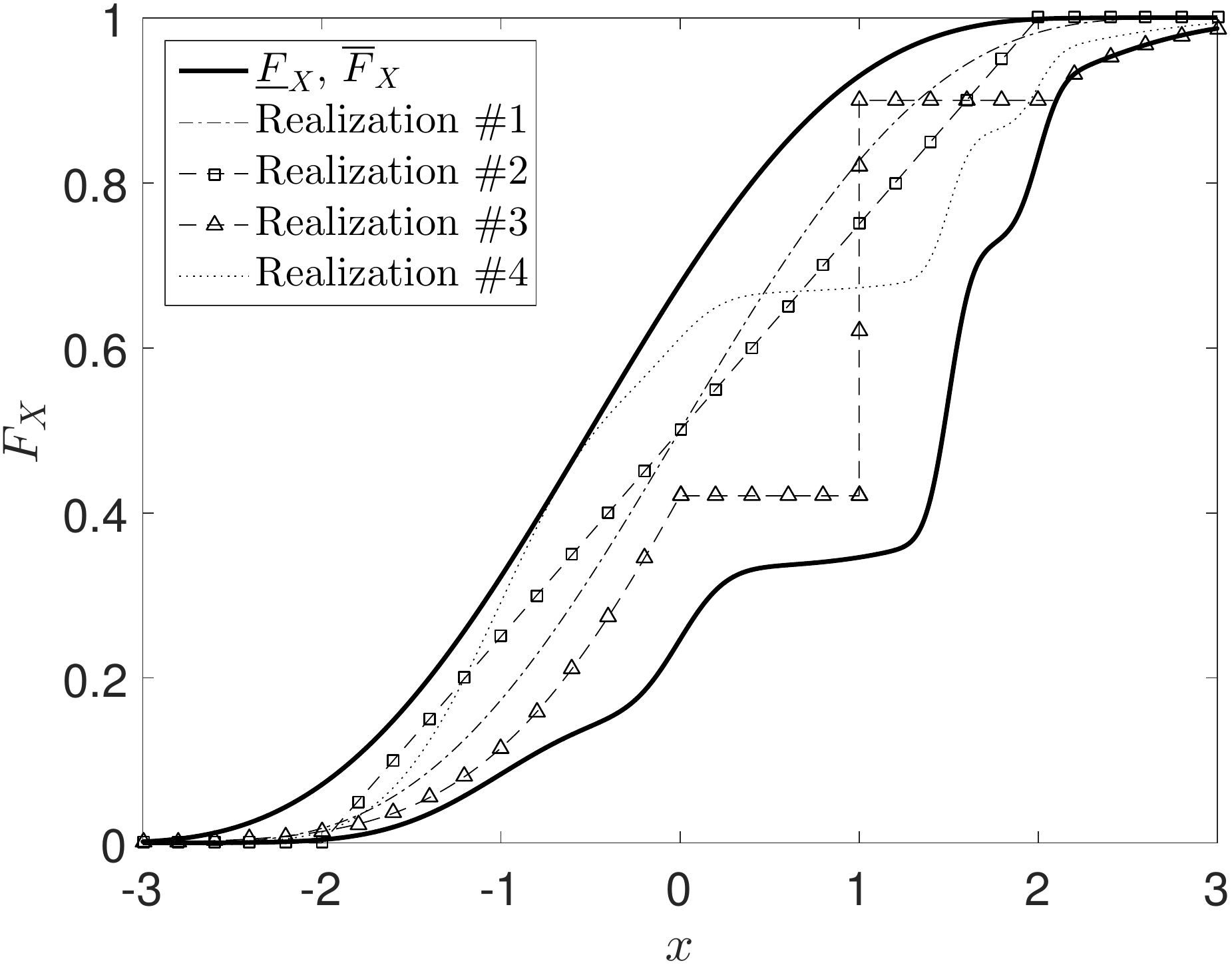}
\caption{{Free probability-box -- boundary curves and realizations of the true but unknown CDF} \label{fig:pbox}}
\end{figure}

The name probability-\emph{box} comes from the fact that $\lF$ and $\uF$ define an intermediate space which resembles a box, as it can be seen in Figure~\ref{fig:pbox}. When the bounds of the p-box coincide for every $x\in X$, \ie $\lF_X(x)=\uF_X(x)$, the corresponding p-box degenerates into a single CDF, as it is usual in standard probability theory. 

The framework of probability-boxes is strongly connected to Dempster-Shafer's theory of evidence \citep{Dempster1967,Shafer1976}, as previously discussed in \eg \cite{Walley2000,Ferson2004a}. The bounds $\lF_X$ and $\uF_X$ can be interpreted as belief and plausibility measures for the event $\{X\leq x\}$ (see also Figure~\ref{fig:pbox}). The \emph{belief} describes the minimum amount of probability that \emph{must} be associated with the event $\{X\leq x\}$ whereas the \emph{plausibility} describes the maximum amount of probability that \emph{might} be associated to the same event. 


\subsection{Determination of p-box bounds}
A number of methods exist for determining the boundary curves of the p-box depending on the type of information accessible \citep{Ferson2003}. These methods include robust Bayesian analysis \citep{Berger1985,Zhang2013}, Chebyshev's inequalities \citep{Oberguggenberger2008,Chebyshev1874}, and Kolmogorov-Smirnov confidence limits \citep{Kolmogoroff1941,Smirnov1939,Zhang2013}. Additionally, the type of information can be diverse: precise or imprecise data from measurements, expert opinions obtained in a survey, or a mixture of both. Hence in the sequel, two cases are distinguished for defining a p-box by aggregation of data in the main part of this paper. These cases lead us to define two scenarios inspired by the engineering practice.

\subsubsection{Case \#1 -- interval-valued expert opinions} \label{sec:pbox:case1}
\begin{figure}[!ht]
\centering
\subfigure[Expert intervals \label{fig:case1:1}]{
	\includegraphics[width = 0.4\linewidth]{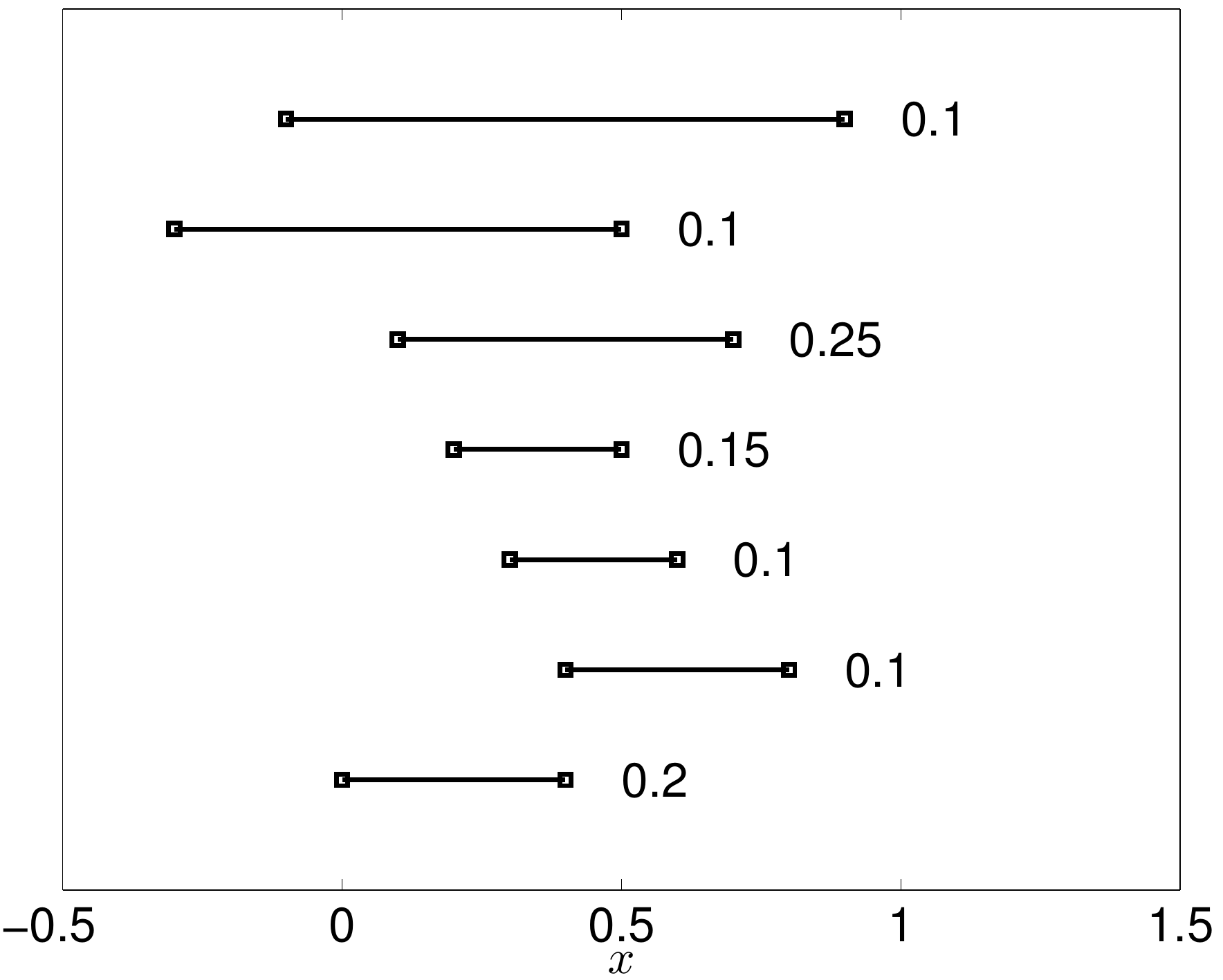}
}
\subfigure[Aggregated p-box \label{fig:case1:2}]{
	\includegraphics[width = 0.4\linewidth]{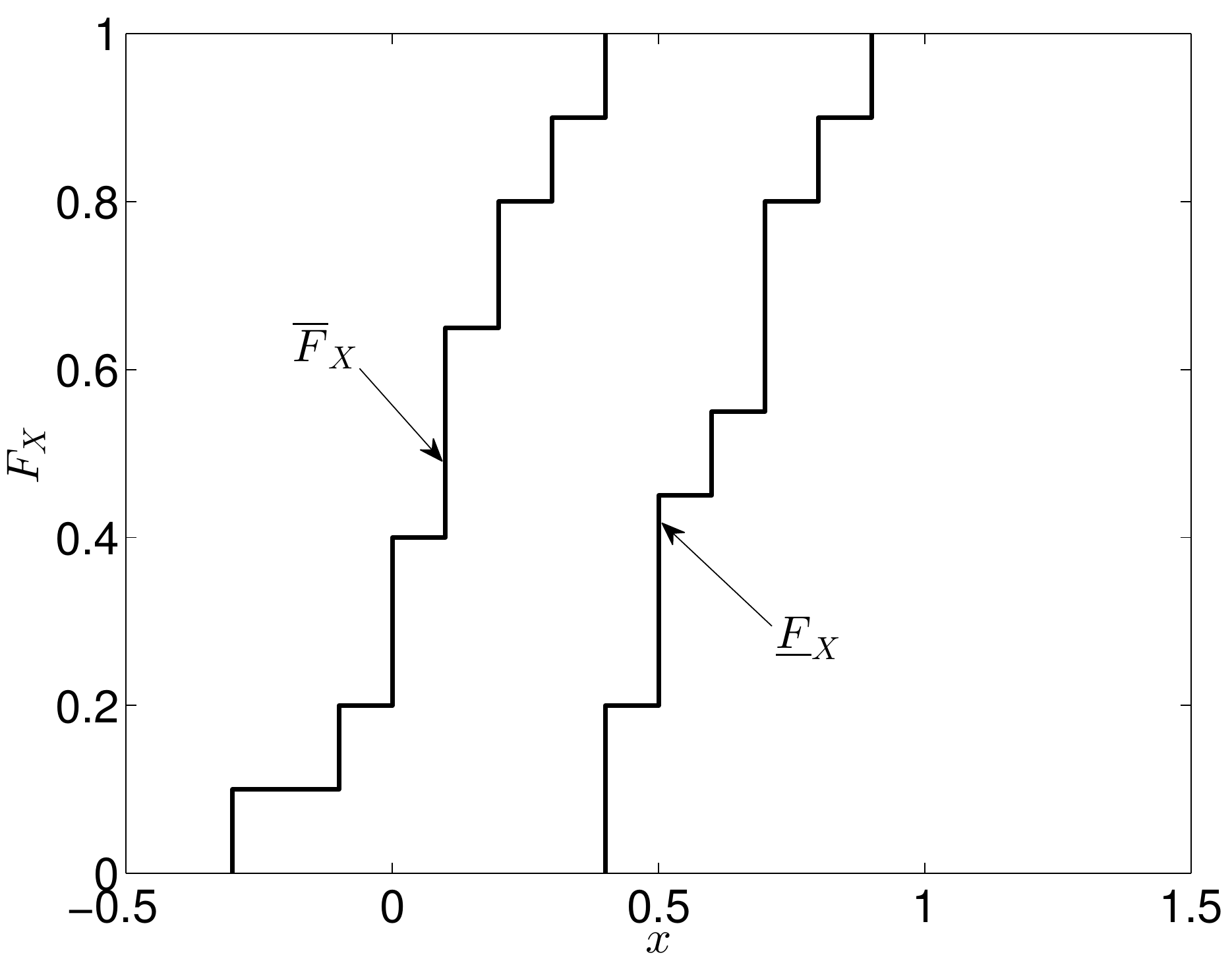}
}
\caption{Case \#1 -- p-box as a combination of expert intervals and credibility measures \label{fig:case1}}
\end{figure}

%
Consider the case where different experts are asked to name an interval for describing the possible values of a variable. Expert $i=1,\ldots,n_E$ provides an interval $x^{(i)}\in\bra{\underline{x}^{(i)},\overline{x}^{(i)}}$. Additionally, a mass of credibility $w^{(i)}$ is assigned to each expert accounting for the expert's knowledge. Note that the credibility is defined here as a relative value so that $\sum_{i=1}^{n_E} w^{(i)} = 1$. Figure~\ref{fig:case1:1} displays a set of seven expert opinions (horizontal intervals) and their credibility (number next to intervals).

A large variety of methods for aggregating multiple sources of information have been proposed in the literature \citep{Ferson2003,Ayyub2001,Ayyub2006}. In this paper, the \emph{mixture method} described in \cite{Ferson2003} is applied under the assumption that the disagreement between the various estimates of the variable represents actual variability (\ie aleatory uncertainty). According to the mixture method, the lower and upper boundary of the p-box are defined as the CDF of the lower bounds $\underline{x}^{(i)}$ and the upper bounds $\overline{x}^{(i)}$ of the intervals taking into consideration their weights:
\begin{equation} \label{eq:case1}
\underline{F}_X(x) = \sum_{i=1}^{n_E} w^{(i)}\cdot \mathbb{I}_{\underline{x}^{(i)}\leq x}(x), \qquad \overline{F}_X(x) = \sum_{i=1}^{n_E} w^{(i)}\cdot \mathbb{I}_{\overline{x}^{(i)}\leq x}(x),
\end{equation}
where $\mathbb{I}_{(\cdot)}$ is the indicator function with $\mathbb{I} = 1$ for a true subscript statement and $\mathbb{I} = 0$ otherwise. The resulting p-box corresponding to the expert opinions in Fig.~\ref{fig:case1:1} is shown in Fig.~\ref{fig:case1:2}.

\subsubsection{Case \#2 -- CDF-shaped expert opinions} \label{sec:pbox:case2}
\begin{figure}[!ht]
\centering
\subfigure[Expert CDFs \label{fig:case2:1}]{
	\includegraphics[width = 0.4\linewidth]{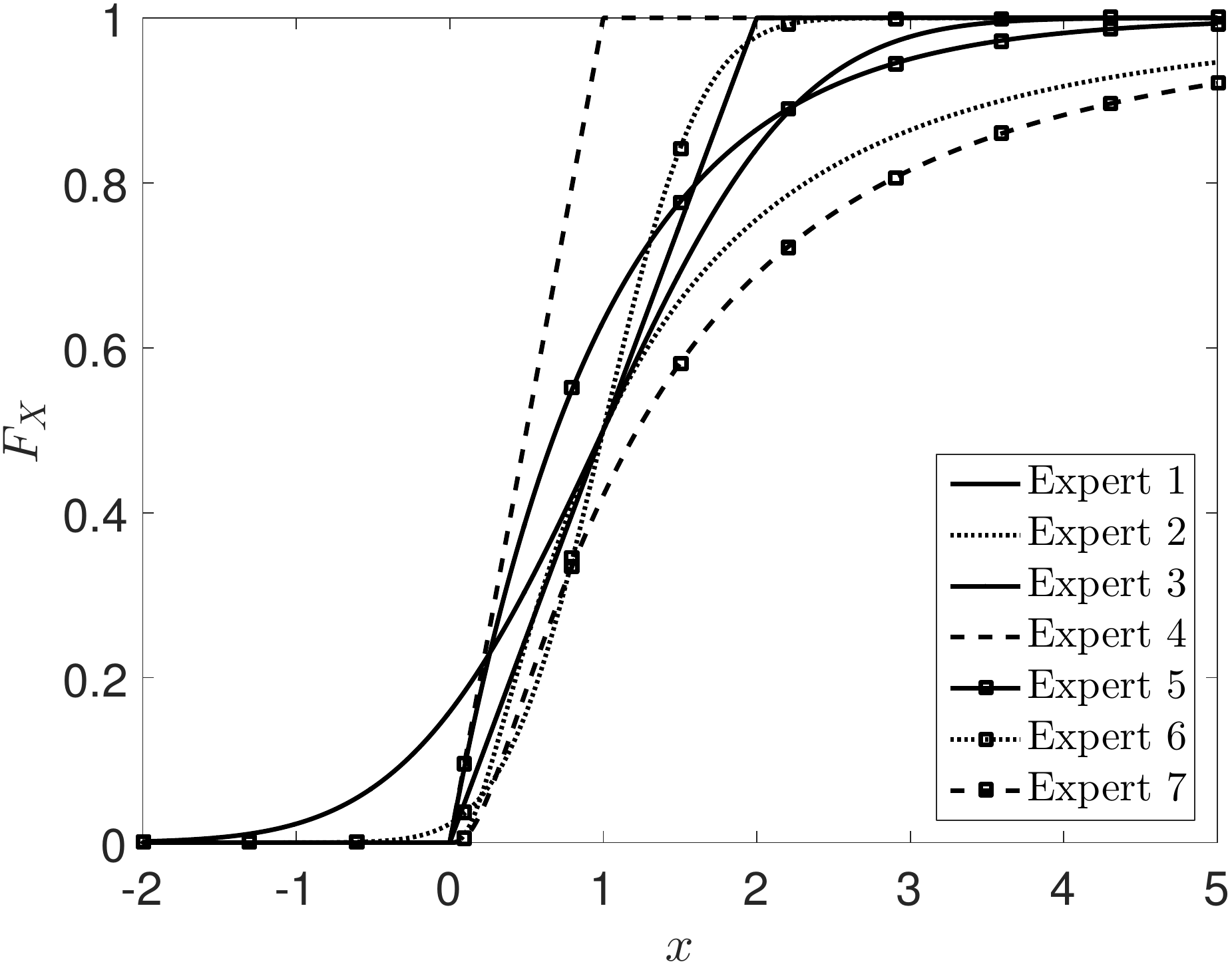}
}
\subfigure[Aggregated p-box \label{fig:case2:2}]{
	\includegraphics[width = 0.4\linewidth]{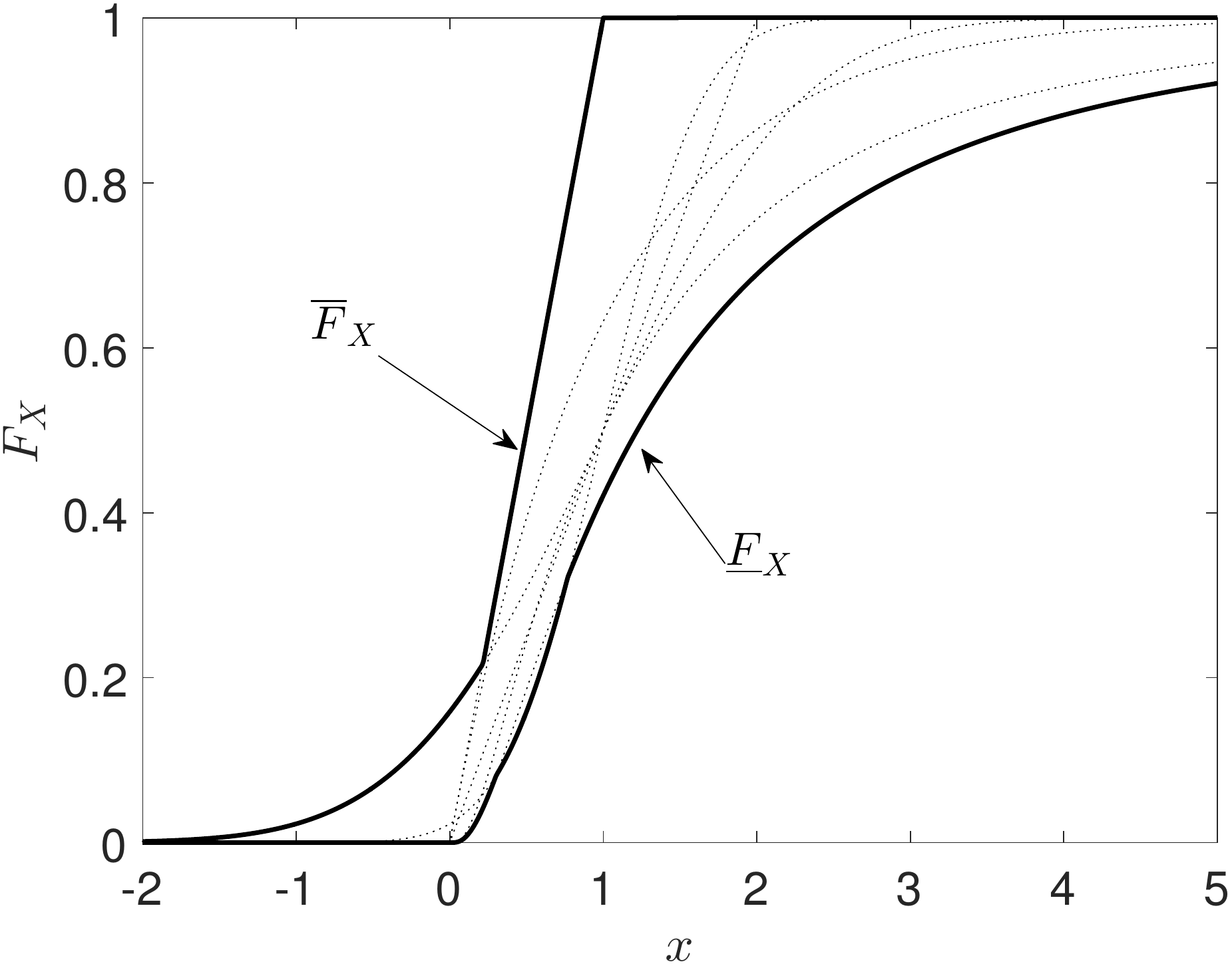}
}
\caption{Case \#2 -- p-box as a combination of expert CDFs \label{fig:case2}}
\end{figure}

%
\paragraph{Case \#2(a)}
Experts are asked to give their opinion on the behaviour of a variable in the form of a CDF. Each expert provides a formulation for the CDF denoted by $F_X^{(i)}, i=1,\ldots,n_E$, where the support of $X$ is possibly unbounded. An example of seven expert CDFs is shown in Figure~\ref{fig:case2:1}. Assuming that these seven CDFs describe the uncertainty in the system (epistemic and aleatory) the p-box can be generated by the envelope of the experts' CDFs \citep{Ferson2003,Fu2011}:
\begin{equation}
\underline{F}_X(x) = \min_{i=1,\ldots,n_E} F_X^{(i)}(x), \qquad \overline{F}_X(x) = \max_{i=1,\ldots,n_E} F_X^{(i)}(x), \qquad \forall x\in\cD_X. 
\end{equation}  
Note that contrary to Case \#1, the credibility of the experts is not considered in Case \#2. Assuming that the true CDF lies between the experts' CDF, the envelope of all opinions includes the true CDF, thus forming a valid p-box. The resulting p-box for the CDFs in Fig.~\ref{fig:case2:1} can be found in Fig.~\ref{fig:case2:2}. The boundary curves of the p-box consist of sections of different input CDFs.

\paragraph{Case \#2(b)}
An alternative way to define a p-box is to choose a distribution family and set interval-valued distribution parameters. The p-box is defined by the envelope of the set of distributions as in Case~\#2(a). The p-box bounds are then found by:
\begin{equation} \label{eq:3}
\lF_X(x) = \min_{\vt\in\mathcal{D}_{\vT}} F_X(x|\vt), \quad \uF_X(x) = \max_{\vt\in\mathcal{D}_{\vT}} F_X(x|\vt), \quad \forall x\in\mathcal{D}_X,
\end{equation}
where $\vt$ is a vector of distribution parameters which are defined in a domain $\mathcal{D}_{\vT}$ \citep{Ferson2003,Zhang2010}. Note that in this case, only the bounds are used for further analyses. The information about the distribution family is ignored (as opposed to so-called parametric p-boxes) once $\lF_X$ and $\uF_X$ are defined by Eq.~(\ref{eq:3}).

\section{Propagation of p-boxes} \label{sec:free}
\subsection{Computational model}

A \emph{computational model} is defined as a deterministic mapping of the $M$-dimensional input vector $\vx$ to the QoI $y$ (considered scalar here for the sake of simplicity):
\begin{equation}
\cm: \vx\in\cD_X\subset\Rr^M \rightarrow y=\cm(\vx)\in\Rr,
\end{equation}
where $\vx = \prt{ x_1,\ldots,x_M }\tr$. The computational model is considered a \emph{black box} of which only the input vector $\vx$ and the QoI $y$ are accessible, as it is usual for legacy computer codes that cannot be modified internally for the sake of uncertainty quantification. Due to uncertainties in the input vector $\vx$, the latter is represented by an imprecise random vector $\vX=\prt{ X_1,\ldots,X_M }\tr$ whose components are assumed statistically independent throughout this paper. Each component $X_i$ is modelled by a p-box, which is propagated through the computational model to an output p-box $Y$:
\begin{equation}
Y = \cm\prt{\vX}.
\end{equation}
In the context of p-boxes, the uncertainty propagation boils down to the estimation of the bounds on the CDF of the QoI $Y$, namely $\underline{F}_Y$ and $\overline{F}_Y$. In the following sections, new algorithms for propagating p-boxes are proposed and discussed.

\subsection{Slicing algorithm} \label{sec:free:slice}

\begin{figure}[ht!]
\centering
\subfigure[Input p-box \label{fig:slicing:1}]{
	\includegraphics[width=0.4\linewidth]{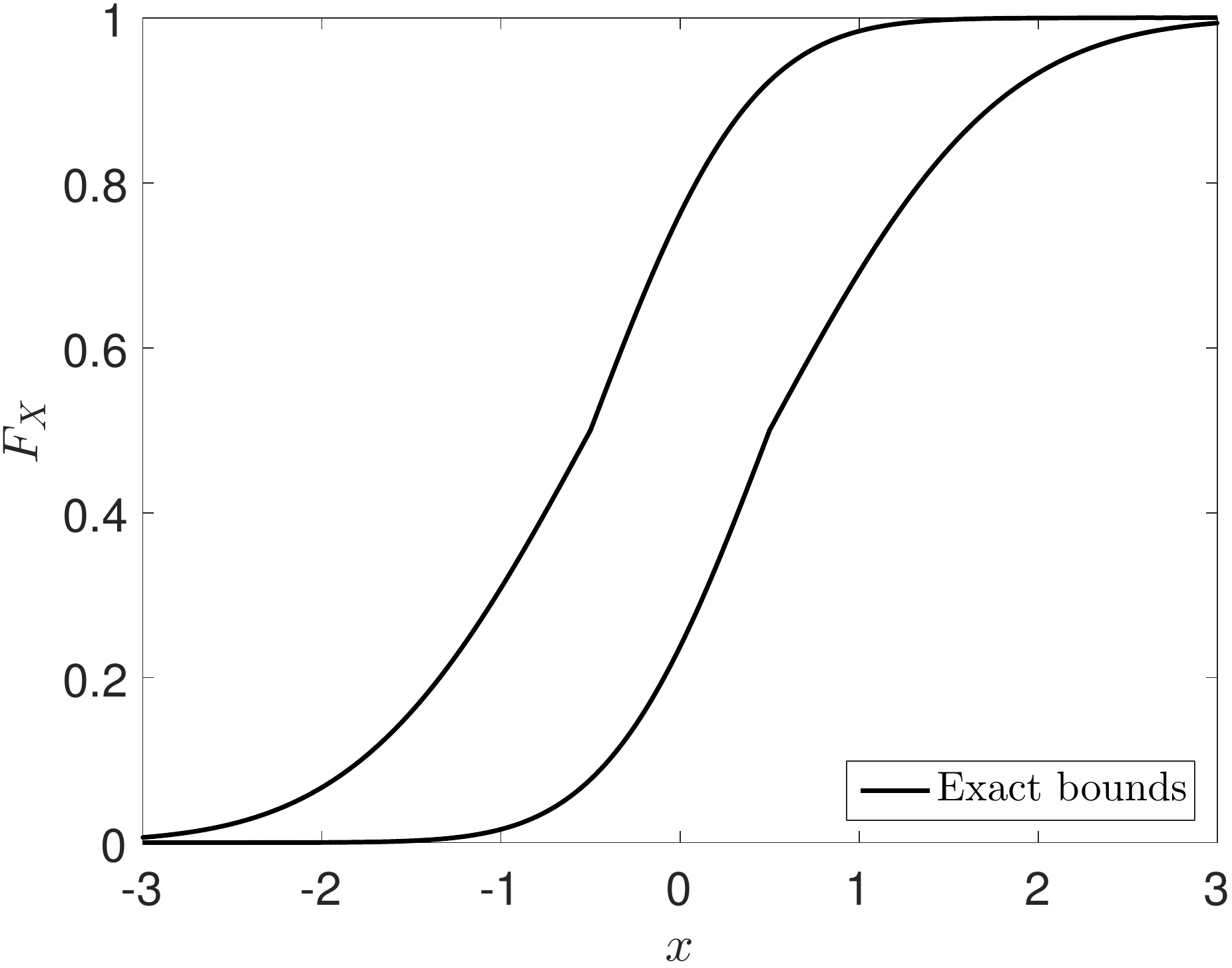}
}
\subfigure[Discretization ($n_X=20$) \label{fig:slicing:2}]{
	\includegraphics[width=0.4\linewidth]{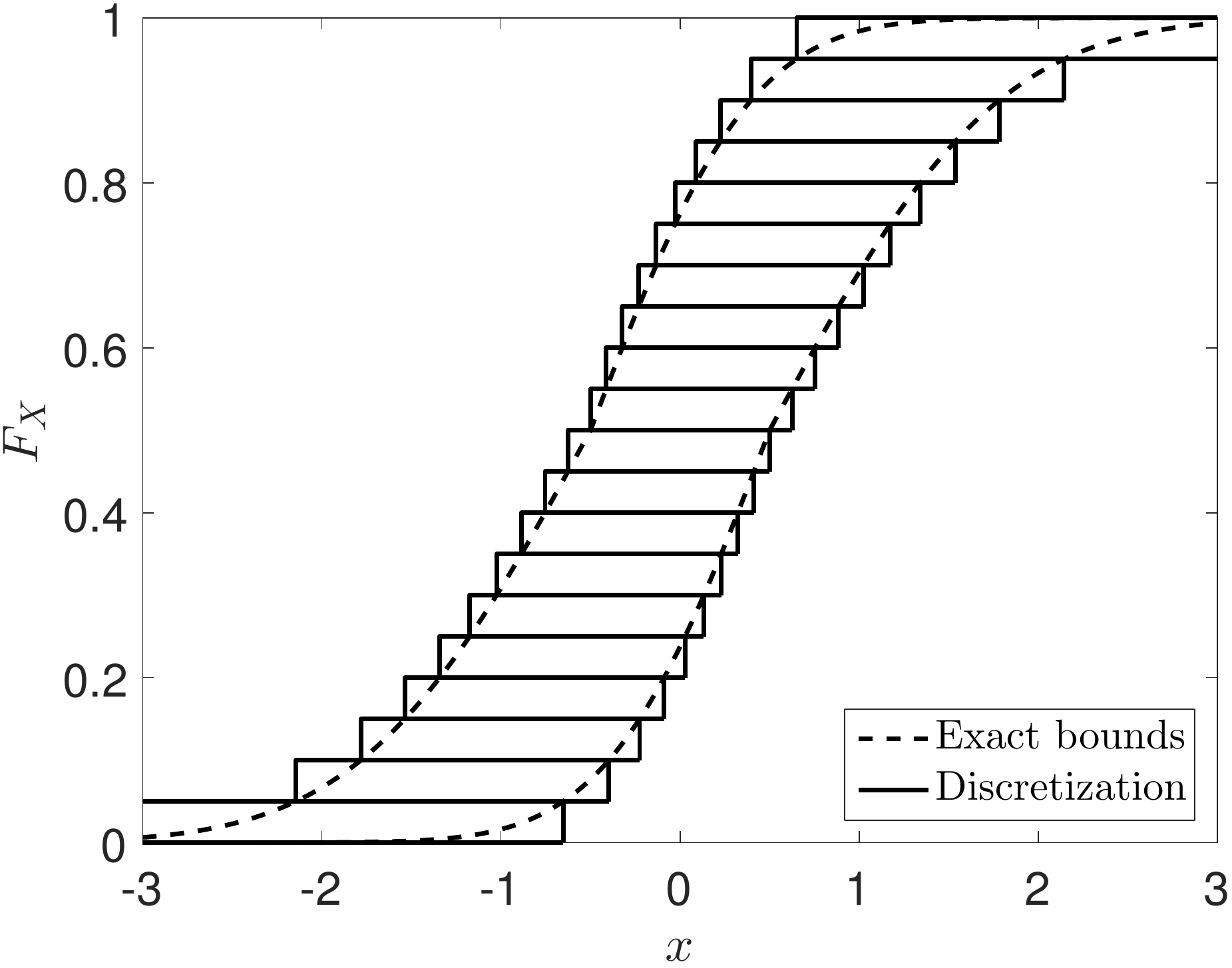}
}\\
\subfigure[Interval propagation ($y=\cm(x)=x/2+4$) \label{fig:slicing:3}]{
	\includegraphics[width=0.4\linewidth]{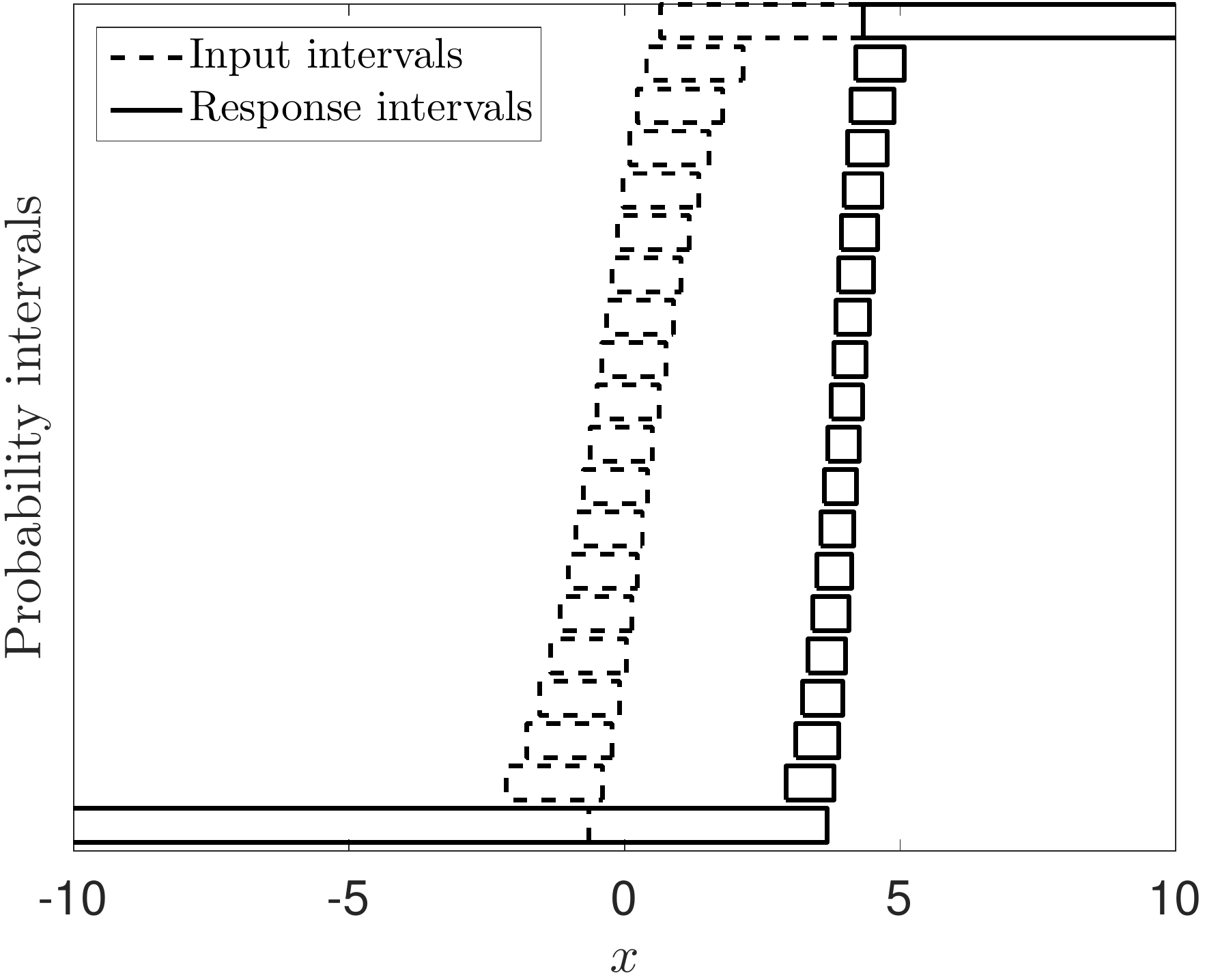}
}
\subfigure[Merging \label{fig:slicing:4}]{
	\includegraphics[width=0.4\linewidth]{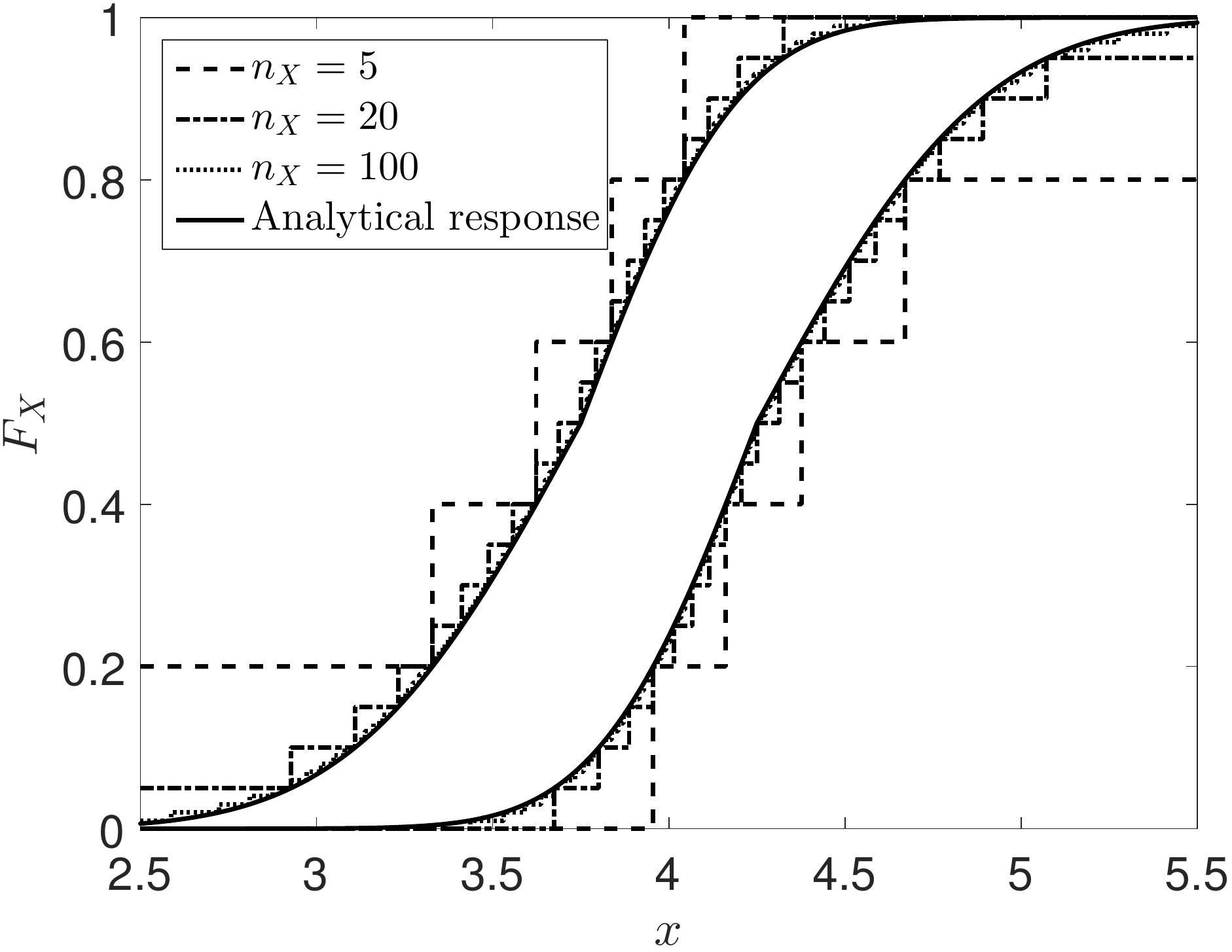}
}
\caption{Illustration of the slicing algorithm. The bounds of the input p-box are defined from Gaussian CDFs with interval-valued mean value and standard deviation. $n_X=20$ intervals are used.  \label{fig:slicing}}
\end{figure}

The \emph{slicing algorithm} transforms the propagation of p-boxes into the propagation of a large number of intervals, the propagation of which is a well-established field of research  related to constraint optimization algorithms \citep{Moore1966,Stolfi2003,Dong1987}. The main steps for applying this algorithm are described in the following: 
\begin{enumerate}
\item  \emph{Discretization}: Discretization methods approximate the p-box by a set of intervals and corresponding probability masses in order to facilitate the uncertainty propagation task. A number of discretization schemes are available in the literature \citep{Tonon2004,Zhang2010}. The \emph{outer discretization method} is now briefly reviewed here. Each input p-box is discretized into a number of interval and associated probability masses. For variable $X_i$, the interval $[0,1]$ is divided into $n_{X_i}$ subintervals with corresponding thickness $m_{i}^{(j)}$ where $j=1,\ldots,n_{X_i}$  and $\sum_j m_{i}^{(j)}=1$ (Figure~\ref{fig:slicing:2}). Let us denote the lower and upper boundary of these intervals by $\underline{c}_{i}^{(j)}$ and $\overline{c}_{i}^{(j)}$, respectively. Given the bounds of the p-box $[\lF_{X_i},\uF_{X_i}]$, the corresponding intervals in $X_i$ are:
\begin{equation} \label{eq:xinterval}
\underline{x}_{i}^{(j)} = \uF_{X_i}^{-1} \prt{ \underline{c}_{i}^{(j)} }, \qquad \overline{x}_{i}^{(j)} = \lF_{X_i}^{-1} \prt{ \overline{c}_{i}^{(j)} },
\end{equation}  
for $j=1,\ldots,n_{X_i}$ and $i = 1,\ldots,M$. The intervals of interest are then $\bra{\underline{x}_{i}^{(j)},\overline{x}_{i}^{(j)}}$ and the associated probability masses are $m_{X_i}^{(j)}$, which together characterize the p-boxes of the input $X_i$. 
\item \emph{Interval propagation}: Let $\cK$ be a set of multi-indices defining a combination of intervals of each input parameter $X_i$:
\begin{equation}
\cK=\acc{\vk=(k_1,\ldots,k_M), \ k_i\in[1,\ldots,n_{X_i}, \ i=1,\ldots,M]  }. 
\end{equation}
Let $\mathcal{D}_{\vk}$ be the hyperrectangle defined by:
\begin{equation}
\mathcal{D}_{\vk} = \bra{\underline{x}_1^{(k_1)},\overline{x}_1^{(k_1)} } \times \ldots \times \bra{\underline{x}_M^{(k_M)},\overline{x}_M^{(k_M)} }.
\end{equation}
For each hyperrectangle $\cD_{\vk}$, two optimization problems are solved to define the associated bounds of the QoI:
\begin{equation} \label{eq:ymin}
\underline{y}^{(\vk)} = \min_{\vx \in\cD_{\vk}} \cm\prt{\vx}, \qquad \overline{y}^{(\vk)} = \max_{\vx \in\cD_{\vk}} \cm\prt{\vx}.
\end{equation}
The probability mass associated to $\cD_{\vk}$ can be computed by:
\begin{equation} \label{eq:m}
m^{(\vk)}_Y = m_{X_1}^{(k_1)} \cdot m_{X_2}^{(k_2)} \cdot\ldots\cdot m_{X_M}^{(k_M)}.
\end{equation}
Correspondingly, the p-box of the QoI is eventually characterized by $n_Y = n_{X_1}\cdot n_{X_2}\cdot\ldots\cdot n_{X_M}$ intervals $\bra{\underline{y}^{(\vk)},\overline{y}^{(\vk)}}$ with associated probability masses given in Eq.~(\ref{eq:m}). Hence, $2\cdot n_Y$ optimization algorithms (see Eq.~(\ref{eq:ymin})) are required in the $M$-dimensional optimization domain in order to propagate the input p-boxes. When $M$ and $n_{X_i}$ become large, this quickly becomes intractable due to the large number of optimizations. This problem is often referred to as the \emph{curse of dimensionality}. 

A number of methodologies can be found in the literature to simplify the optimizations, amongst which are the classical interval analysis \citep{Moore1966}, affine arithmetic \citep{Stolfi2003} and the vertex method \citep{Dong1987,Dubois2004}. However, these simplifications require restrictive assumptions, such as monotonicity in the computational model, to ensure accuracy. Other optimization algorithms can be applied such as local, derivative-based methods (\eg BFGS algorithm \citep{Byrd1999}), global methods (\eg genetic algorithms \citep{Goldberg1989} and differential evolution algorithms \citep{Storn1997,Deng2013}) and hybrid methods (\ie with a global and local component) for solving directly Eq.~(\ref{eq:ymin}). They are more accurate in the general case but require extensive computational resources. 
\item \emph{Merging}: The result of the previous step is a set of $n_Y=|\cK|$ intervals $\bra{\underline{y}^{(\vk)},\overline{y}^{(\vk)}}$ and the corresponding probability masses $m_y^{(\vk)}$. The response p-box is then obtained by converting $\acc{\overline{y}^{(\vk)},m_y^{(\vk)}}$ and $\acc{\underline{y}^{(\vk)},m_y^{(\vk)}}$ to weighted empirical CDFs $\lF_Y$ and $\uF_Y$ as explained in Eq.~(\ref{eq:case1}).
\end{enumerate}


Figure~\ref{fig:slicing} illustrates the main steps of the slicing algorithm on a one-dimensional problem. The bounds of the input p-box are defined by two Gaussian distributions with $\mu_X = \bra{1.5,\, 2}$ and $\sigma_X = [0.7,\ 1.0]$ (Case~\#2(b)) (Figure~\ref{fig:slicing:1}). The p-box is discretized with $n_X=20$ equally spaced subintervals, \ie $m^{(j)}=1/20$ (Figure~\ref{fig:slicing:2}). The computational model is $y = x/2+4$ (Figure~\ref{fig:slicing:3}). Figure~\ref{fig:slicing:4} illustrates the influence of $n_X$ on the accuracy of the response p-box. It can be seen from this simple example already that the number of discretization points $n_{X_i}$ is crucial to the accuracy of the response p-box: the approximated response p-box is conservative in the sense that it is wider than the exact response p-box obtained analytically. The larger $n_{X_i}$ the more accurate the approximated p-box, and at the same time the larger the computational costs. This effect is more pronounced when the input vector is multi-dimensional, as discussed later in Section~\ref{sec:appl:nd}.  

Note that in Case \#1, the intervals might be chosen to represent each expert's opinion. Then, the expert's credibility is equal to the probability mass in uncertainty propagation, \ie $m^{(j)} = w^{(j)}, \ j=1,\ldots,n_{X_i}$. 

\subsection{Problem conversion} \label{sec:free:problem}

A disadvantage of the slicing algorithm is the full-factorial design, which leads to an {exponentially} large number of optimizations for high-dimensional problems, namely $n_Y=|\cK|=n_{X_1}\cdot n_{X_2}\cdot \ldots\cdot n_{X_M}$. In order to circumvent the effects of the full factorial approach, the imprecise problem setting is reformulated in this section, as originally proposed by \cite{Zhang2010,SchoebiESREL2015}. {They replace the full factorial design of the slicing algorithm with a random sampling-based approach as follows. Consider the slicing algorithm with a large number of subintervals $n_{X_i}\rightarrow\infty$ (and $|\cK|\rightarrow\infty$) in order to accurately estimate the response p-box $Y$. The corresponding number of response intervals will be $n_Y\rightarrow\infty$, too.  Then, $Y$ can be approximated by propagating a random subset of $\cK'\subset\cK$ instead of the entire set $\cK$ to reduce the computational costs. This approximation allows to estimate $Y$ more efficiently than the full-factorial design.} 

{Instead of creating an infinite set $\cK$, the same problem can be interpreted with probabilistic variables.} Consider the random vector $\vC$ made of independent, uniformly distributed variables in the unit-hypercube domain $\cD_{\vC} = [0,1]^M$. The random variable $C_i$ shall describe the CDF value of input variable $X_i$, \ie $C_i = F_{X_i}(X_i)$. {In other words, $C_i$ describes the index of one of the $n_{X_i}\rightarrow\infty$ subintervals in $\cK$.} Given a p-box in $X_i$, each $c_i\in [0,1]$ corresponds then to an interval $\bra{\underline{x}_i,\overline{x}_i}$ in $\cD_{X_i}$ through the inverse CDF of the bounds of the p-box:
\begin{equation} \label{eq:xsample}
\underline{x}_i(c_i) = \uF^{-1}_{X_i}(c_i), \qquad \overline{x}_i(c_i) = \lF^{-1}_{X_i}(c_i).
\end{equation}
Note that compared to the previous definition in Eq.~(\ref{eq:xinterval}) where the interval $\bra{\underline{c}_i,\overline{c}_i}$ is the input argument, Eq.~(\ref{eq:xsample}) uses a single value $c_i$. Eq.~(\ref{eq:xsample}) can be interpreted from Eq.~(\ref{eq:xinterval}) when setting $n_{X_i}\rightarrow \infty$ and thus $m^{(j)}_i\rightarrow0$ and $\underline{c}_i\approx\overline{c}_i$. For a given realization $\vc\in\cD_{\vC}$, let us denote by: $\cD_{\vc}=\bra{\underline{x}_1(c_1),\overline{x}_1(c_1)}\times \ldots \times \bra{\underline{x}_M(c_M),\overline{x}_M(c_M)}$. As a consequence, two computational models can then be formulated as a function of $\vc$, equivalent to Eq.~(\ref{eq:ymin}) \citep{SchoebiESREL2015,AlvarezICVRAM2014}:
\begin{equation} \label{eq:lcm}
\underline{y}=\lcm(\vc) = \min_{\vx \in \cD_{\vc}} \cm(\vx), \qquad \overline{y}=\ucm(\vc) =  \max_{\vx \in \cD_{\vc}} \cm(\vx).
\end{equation}
These equations lead to intervals $\bra{\underline{y},\overline{y}}$ of the QoI $Y$. In analogy with the slicing algorithm (see also Figure~\ref{fig:slicing}), the lower bounds of the intervals model the upper boundary curve of $Y$ (and vice versa). Hence, the lower bound model $\lcm$ maps $\vC$ to the upper bound $\overline{Y}$ and the upper bound model $\ucm$ maps $\vC$ to the lower bound response $\underline{Y}$:
\begin{equation} \label{eq:Ymin}
\underline{Y} = \ucm\prt{\vC}, \qquad \overline{Y} = \lcm\prt{\vC},
\end{equation}
where $\underline{Y}$ and $\overline{Y}$ are characterized by the CDF $\lF_Y$ and $\uF_Y$ respectively. Eq.~(\ref{eq:Ymin}) splits the propagation of p-boxes into two standard uncertainty propagation problems associated with input random vector $\vC$. 

The probabilistic description of the auxiliary input vector allows for conventional methods of uncertainty propagation such as random sampling (Monte Carlo simulation, Latin-hypercube sampling \citep{McKay1979}) and low-discrepancy sequences (Sobol' sequence \citep{sobol1967}, Halton sequence \citep{Halton1960}). These methods are more efficient than the full factorial design approach in Section~\ref{sec:free:slice}, but they are not conservative with respect to the p-box due to the nature of sampling methods, as opposed to the previously discussed slicing algorithm. Large sample sets must then be used to ensure proper accuracy.

\section{Two-level meta-modelling}\label{sec:2meta}
\subsection{Basic idea} \label{sec:2meta:idea}

Considering the main steps of the uncertainty propagation of p-boxes presented in Section~\ref{sec:free:problem}, there are three main factors contributing to the total computational effort. Firstly, the computational model is evaluated a large number of times due to the optimization methods, in particular when using global optimization methods in the general case when model monotonicity does not hold. Secondly, the number of optimization operations is large considering the sampling-based approach for estimating the bounds of the response p-box, \ie $\underline{Y}$ and $\overline{Y}$. Last but not least, the cost of a single evaluation of the computational model may affect the total costs considerably. 

In order to address these three factors, it is proposed in this paper to surrogate the computational model at two levels by polynomial chaos expansions. More specifically, the first-level meta-model approximates the response of $\cm$, whereas the second-level one approximates the response of the lower and upper model denoted by $\lcm$ and $\ucm$. 

\subsection{Meta-modelling $\lcm$ and $\ucm$} \label{sec:2meta:pce}

\subsubsection{Polynomial Chaos Expansions}
Assume the output $Y=\cm\prt{\vC}$ of a computational model $\cm$ with input random vector $\vC\sim f_{\vC}$ has a finite second moment, \ie $\mathbb{E}\bra{Y^2}<+\infty$. According to \citep{GhanemBook2003}, $Y$ may be cast as the following polynomial chaos expansion:
\begin{equation} \label{eq:pce0}
Y= \cm\prt{\vC} = \sum_{\ua\in\mathbb{N}^M} a_{\ua} \psi_{\ua}\prt{\vC},
\end{equation} 
where $\acc{\psi_{\ua}(\vC), \ \ua\in\mathbb{N}^M}$ are multivariate orthonormal polynomials in $\vC$ and $a_{\ua}$ are coefficients to be computed.
Due to the assumption of independence, the joint CDF of $\vC$ is the product of its marginals. Then, for each marginal distribution $f_{C_i}$ a functional inner product is defined as: 
\begin{equation}
\langle \phi_1,\phi_2\rangle_i = \int_{\cD_{i}} \phi_1(c_i)\, \phi_2(c_i) \, f_{C_i}(c_i)\,\text{d}c. 
\end{equation} 
Then, a family of orthonormal polynomials $\acc{\psi_j^{(i)}, \, j\in\mathbb{N}}$ can be built for each input variable $C_i$ that satisfies:
\begin{equation}
\langle \psi_j^{(i)}, \psi_k^{(i)} \rangle = \int_{\cD_{C_i}} \psi_j^{(i)}(c) \, \psi_k^{(i)}(c)\, f_{C_i}(c_i) \, \text{d}c = \delta_{jk},
\end{equation}
where $\delta_{jk} = 1$ for $j=k$ and $\delta_{jk} = 0$ otherwise. In the present case of uniform random variables in $C_i$, Legendre polynomials forms the family of orthonormal polynomials. For other orthogonal polynomial function families, the reader is referred to \eg {\cite{SudretHDR,Xiu2002}}. The multivariate polynomials (used in Eq.~(\ref{eq:pce0})) are finally composed of the univariate polynomials by tensor product:
\begin{equation}
\psi_{\ua}(\vC)=\prod_{i=1}^{M} \psi_{\alpha_i}^{(i)}\prt{C_i}.
\end{equation}

\subsubsection{Non-intrusive PCE}

In practice, it is not feasible to handle infinite series as presented in Eq.~(\ref{eq:pce0}). Hence, the infinite set of multivariate orthonormal polynomials is truncated, such that Eq.~(\ref{eq:pce0}) transforms to:
\begin{equation} \label{eq:pcetrunc}
Y\approx Y^{(P)} \eqdef \cm^{(P)}(\vC) = \sum_{\ua\in\cA} \va_{\ua} \psi_{\ua}(\vC),
\end{equation}
where $\cA\subset \mathbb{N}^M$ is a finite set of multi-indices and $P$ denotes its cardinality ($P=|\cA|$). A number of truncation schemes have been proposed in the literature, amongst which is the \emph{hyperbolic truncation set} \cite{BlatmanPEM2010}, which is based on the q-norm: 
\begin{equation}
\cA^{M,p}_q = \acc{ \ua\in\mathbb{N}^M: \ ||\ua||_q\leq p},
\end{equation}
\begin{equation}
||\ua||_q = \prt{\sum_{i=1}^{M}\alpha_i^q}^{1/q},
\end{equation}
where $0<q\leq 1$ is a user-defined parameter and $p$ is the maximal total degree of the polynomials. A small value of $q$ leads to a smaller number of high-rank polynomials. When $q$ tends to zero, only univariate polynomials are left in the set of polynomials. Using $\cA^{M,p}_q$ then, Eq.~(\ref{eq:pcetrunc}) transforms to:
\begin{equation} \label{eq:pce}
Y\approx Y^{(P)} \eqdef \cm^{(P)}(\vC) = \sum_{\ua\in\cA_q^{M,p}} a_{\ua} \psi_{\ua}(\vC), \quad P=|\cA_q^{M,p}|.
\end{equation}

Finally, the coefficients $a_{\ua}$ are computed. Assume that the variable $Y$ in Eq.~(\ref{eq:pce}) is replaced by $\underline{Y}$ and $\overline{Y}$ for the two computational models in Eq.~(\ref{eq:lcm}). Due to interpreting $\cm$ and thus $\lcm$ and $\ucm$ as black-box models, only \emph{non-intrusive} training algorithms are presented here to compute the PCE coefficients. One strategy is least-square analysis, as originally introduced by {\cite{Tatang97,Berveiller2006}}. Consider a set of $N$ samples of the input vector $\vCC = \acc{ \vcc^{(1)},\ldots,\vcc^{(N)} }$, called \emph{experimental design}, and the corresponding responses of the two computational models: 
$$\underline{\cY} = \acc{ \underline{\cY}^{(1)} = \ucm\prt{\vcc^{(1)}},\ldots,\underline{\cY}^{(N)} = \ucm\prt{\vcc^{(N)}} },$$ 
$$\overline{\cY} = \acc{ \overline{\cY}^{(1)} = \lcm\prt{\vcc^{(1)}},\ldots,\overline{\cY}^{(N)} = \lcm\prt{\vcc^{(N)}} }.$$ 
The set of coefficients can be computed through the solution of the least squares problem:
\begin{equation}
\overline{\va}^* = \arg\min_{\va\in\Rr^{P}} \frac{1}{N}\sum_{i=1}^{N} \prt{\underline{\cY}^{(i)}-\sum_{\ua\in\cA^{M,p}_q} a_{\ua}\psi_{\ua}\prt{\vcc^{(i)}} }^2,
\end{equation}
\begin{equation}
\underline{\va}^* = \arg\min_{\va\in\Rr^{P}} \frac{1}{N}\sum_{i=1}^{N} \prt{\overline{\cY}^{(i)}-\sum_{\ua\in\cA^{M,p}_q} a_{\ua}\psi_{\ua}\prt{\vcc^{(i)}} }^2.
\end{equation}
Finally, the two PCE models for the bounds of the response p-box are:
\begin{equation}
\underline{Y}\approx \ucm^{(\text{P})}\prt{\vC} = \sum_{\ua\in\cA^{M,p}_q} \overline{a}^*_{\ua} \psi_{\ua}\prt{\vC},
\end{equation}
\begin{equation}
\overline{Y}\approx \lcm^{(\text{P})}\prt{\vC} = \sum_{\ua\in\cA^{M,p}_q} \underline{a}^*_{\ua} \psi_{\ua}\prt{\vC}.
\end{equation}

\subsubsection{Sparse PCE}

The efficiency of the meta-modelling algorithm depends on the choice of the set of polynomials $\cA^{M,p}$ or the hyperbolic truncation set $\cA^{M,p}_q$, which decrease drastically the number of unknowns of the problem while ensuring that polynomials up to degree $p$ in each variable are considered. In high-dimensional problems, however, such truncation schemes are not efficient enough. 

A complementary approach is to select the polynomials out of a candidate set which are most influential to the system response. Different selection algorithms have been presented in the literature, such as \emph{least absolute shrinkage operator} (LASSO) \citep{Tibshirani1996}, \emph{least angle regression} (LAR) \citep{Efron2004,BlatmanJCP2011}, and \emph{compressive sensing} {\citep{Sargsyan2014a,Doostan2011}}. In this paper, sparse PCE meta-models are trained with LAR which has also been applied previously in \cite{SchobiSudretIJ4UQ2015,SchobiASCE2015}.  Note that the resulting  $\cA$ is likely to be different for $\underline{Y}$ and $\overline{Y}$, because of the selection algorithm.

\subsection{Meta-modelling $\cm$} \label{sec:2meta:first}
\subsubsection{Condensation of p-boxes}

The expensive-to-evaluate computational model $\cm$ may itself be approximated with a sparse PCE model in view of solving the global optimization problems defining $\lcm$ and $\ucm$. The input of $\cm$ is a p-box in $M$ dimensions in the problems on consideration. However, in order to apply a sparse PCE model, a probabilistic input vector is required. Hence, auxiliary input variables $\widetilde{X}_i$ are defined for the sole purpose of meta-modelling $\cm$.
The auxiliary input variables should represent the probability mass in the input p-boxes in an appropriate manner. In other words, the auxiliary distributions aim at ''summarizing'' the p-box, such as fulfilling $\lF_X(x)<F_{\widetilde{X}}(x)<\uF_X(x), \, \forall x\in\mathcal{D}_X$.  This is the so-called \emph{condensation} phase. As p-boxes are defined on interval-valued CDFs, a number of distributions can be proposed, \ie there is no unique choice.
 
Generally speaking, for Case \#1 (Section~\ref{sec:pbox:case1}) and bounded p-boxes, it is proposed to use a uniform distribution between the minimum and maximum value of the p-boxes, \ie:
\begin{equation} \label{eq:auxU}
\widetilde{X}_i \sim \mathcal{U}\prt{\min_{j=1\ldots,n_E}\prt{\underline{x}_i^{(j)}},\max_{j=1,\ldots,n_E}\prt{\overline{x}_i^{(j)}}},
\end{equation}
where $\bra{\underline{x}_i^{(j)},\overline{x}_i^{(j)}}$ is the interval on variable $i$ corresponding to the $j$-th expert. For Case \#2 (Section~\ref{sec:pbox:case2}) and generally for unbounded p-boxes, it is proposed to define the CDF of $\widetilde{X}_i$ as an average curve of its input p-boxes:
\begin{equation} \label{eq:tilder}
\widetilde{F}_{X_i}(x_i) = \frac{1}{2} \prt{\lF_{X_i}(x_i)+\uF_{X_i}(x_i)}.
\end{equation}
In both cases, the proposed auxiliary input distribution covers the p-box in the areas where most of the probability mass is located and is therefore suitable to represent the p-box in a sparse PCE model. However, in terms of accuracy of the meta-model, different distributions might be more suitable than the one in Eq.~(\ref{eq:tilder}) (see also Section~\ref{sec:appl} for a detailed discussion on this aspect). 

\subsubsection{Arbitrary input PCE}

Having defined an auxiliary input vector $\widetilde{\vX}$, it can be propagated through the computational model. The corresponding uncertainty propagation problem then reads:
\begin{equation}
\widetilde{Y} = \cm\prt{\widetilde{\vX}}, 
\end{equation}
and the corresponding approximation with PCE:
\begin{equation}
\widetilde{Y}\approx \cm^{(\text{P})}(\widetilde{\vX}) = \sum_{\ua\in\cA^{M,p}_q} \widetilde{a}_{\ua} \psi_{\ua}\prt{\widetilde{\vX}}.
\end{equation}
This PCE model can be trained by pure vanilla least-square analysis, least-angle regression or any other non-intrusive technique (see Section~\ref{sec:2meta:pce}).

An important aspect here is when $\widetilde{X}_i$ has an arbitrary CDF shape for which it might not be trivial to define a set of orthogonal polynomials \citep{Gautschi2004}. \cite{Ahlfeld2016,Dey2016} discuss the use of so-called \emph{arbitrary PCE}, which are based on arbitrarily shaped input distributions. An alternative way is to formalize an isoprobabilistic transform from variables $\widetilde{X}_i$ to $Z_i$ for which a suitable set of orthogonal polynomials is known already. The mapping from one distribution to the other is denoted by $T: Z_i = T\prt{{\widetilde{X}_i}}$. A point $x_i\sim \widetilde{X}_i$ can be mapped to $Z_i$ as $z_i = F_{Z_i}^{-1}\prt{F_{\widetilde{X}_i}\prt{x_i}}$. When $\widetilde{X}_i$ and $Z_i$ are chosen in a smart way, the transform $T$ shall be nearly linear. The PCE model can then be written as:
\begin{equation}
\widetilde{Y} \approx \cm^{(\text{P})}(\widetilde{\vX}) = \sum_{\ua\in\cA^{M,p}} \widetilde{a}_{\ua} \psi_{\ua}\prt{T\prt{\widetilde{\vX}}},
\end{equation}
where the set of $\psi_{\ua}$'s are orthogonal with respect to the vector $\vZ=T\prt{\widetilde{\vX}}$.


\subsection{Aggregation of two levels of meta-modelling} \label{sec:free:discussion}
\subsubsection{Framework of meta-models}

\begin{figure}[ht!]
\centering
\includegraphics[width=0.7\linewidth]{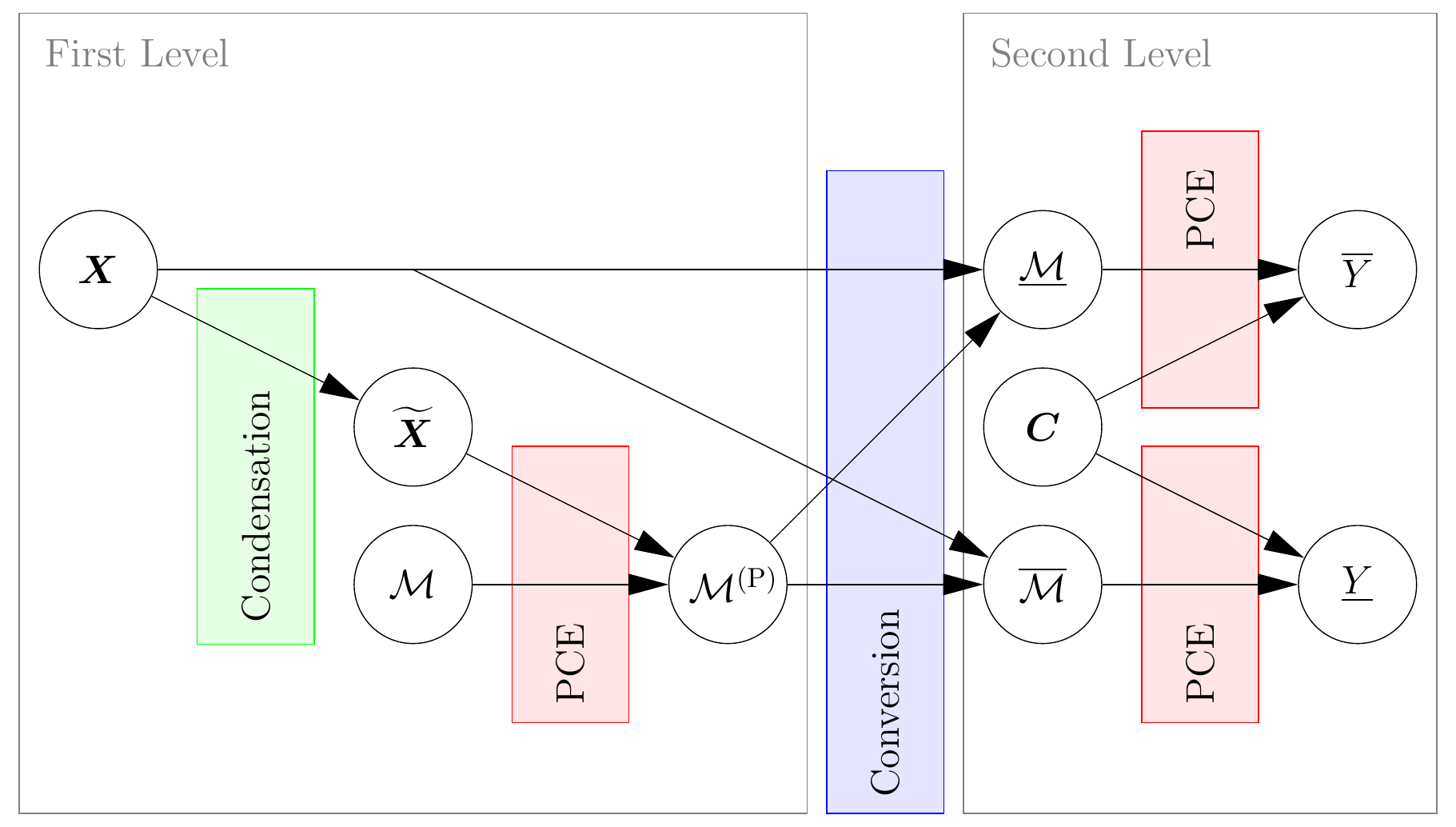}
\caption{Two-level meta-modelling for the propagation of p-boxes -- definition and connection of the two levels \label{fig:2meta}}
\end{figure}

Figure~\ref{fig:2meta} summarizes the two levels of meta-modelling presented previously. A first level of sparse PCE approximates the computational model $\cm$ on the basis of the auxiliary input vector $\widetilde{\vX}$, which is itself defined according to the type of input p-box considered. This results in the meta-model $\cm^{(\text{P})}$. Then, the problem is divided into the estimation of $\underline{Y}$ and $\overline{Y}$ by Eq.~(\ref{eq:Ymin}). They are obtained by approximating $\ucm$ and $\lcm$, respectively, via sparse PCE and the probabilistic input vector $\vC$, which defines the second level of the two-level approach.  

\subsubsection{Case \#1}

In the case of bounded p-boxes, \ie when the support of the boundary CDFs $\lF_{X_i}$ and $\uF_{X_i}$ are compact, the methods can be applied as described in the previous sections. For the first level of meta-modelling, bounded variables $\widetilde{\vX}$ are used so that the response $\widetilde{Y}$ may be approximated efficiently. On the second-level meta-model however, the response $\underline{Y} = \ucm(\vC)$ and $\overline{Y} = \lcm(\vC)$ are non-smooth due to the stairs-like input p-boxes $X_i$ originating in a finite set of expert intervals. The effect of this non-smoothness is investigated further in the example in Section~\ref{sec:appl:2d}. However, when the number of expert intervals and the number of variables are small, the second-level meta-model may be redundant, as the full factorial design approach is tractable (\ie $n_Y$ small). In the other extreme case of large number of variables and intervals, the response boundary curves loose the stairs-shaped nature of the CDF curves due to the large number of response intervals (\ie $n_Y$ large).

\subsubsection{Case \#2}

In the case of unbounded p-boxes, the uncertainty propagation analysis, as proposed above, may become inefficient due to the usage of a bounded $\vC$ on the second-level meta-model. In general, it is not advised to model an unbounded variable $X$ by a bounded variable $Y$ and the corresponding isoprobabilistic transform, due to the highly non-linear functional form of this transform \citep{SchobiAPSSRA2016,SchoebiESREL2015}. 

In order to reduce the effect of the non-linear transform and to ensure fast convergence, the second-level meta-model is trained in the auxiliary input domain $\widetilde{\vX}$ as previously discussed in \cite{SchoebiESREL2015}. An isoprobabilistic transform can be formulated such that $\vC = T\prt{\widetilde{\vX}}$. The corresponding models are:
\begin{equation}
\underline{Y} = \ucm\prt{T\prt{\widetilde{\vX}}}, \qquad \overline{Y} = \lcm\prt{T\prt{\widetilde{\vX}}}. 
\end{equation} 
These two computational models can be approximated using sparse PCE meta-models. 

\subsubsection{Special case: monotone computational models}

When the computational model $\cm$ is known to be monotonic with respect to all variables $X_i$, then the constraint optimization in Eq.~(\ref{eq:lcm}) reduces to the analysis of the corners of the search domain. Hence, uncertainty propagation of p-boxes simplifies to the propagation of the bounds of the input p-boxes. It follows that the meta-modelling of $\lcm$ and $\ucm$ becomes out of scope. $\underline{Y}$ and $\overline{Y}$ can be directly estimated based on $\cm^{(\text{P})}$ and the bounds of each $X_i$. 

\subsubsection{{Convergence behaviour}}
{The two-level meta-modelling algorithm consists of three main components that contribute to its overall performance. Each of those components is a potential source of inaccuracies. The two levels of meta-modelling introduce approximation errors due to the finite size of the experimental design and the use of regression-based meta-modelling techniques. Moreover, the accuracy of the extreme values resulting in the optimization algorithm depends on the optimization algorithm itself. Hence, it is crucial to monitor the accuracy of all three components in an uncertainty quantification analysis in order to ensure convergent results.}

{The following examples highlight the performance of the two-level meta-modelling algorithm on a benchmark analytical function as well as two engineering problem settings.  The general convergence behaviour is analysed empirically by numerically experiments and convergence issues are pointed out.}

\section{Applications} \label{sec:appl}
\subsection{Reference solution}

In the following application examples, the proposed two-level meta-modelling algorithm is compared to a reference solution. The reference solution is obtained by (i) using the original computational model $\cm$ (no first-level meta-model $\cm^{(P)}\prt{\widetilde{\vX}}$) and by (ii) using a large number of points ($n=10^6$) in the prediction of $\underline{Y}$ and $\overline{Y}$.

\subsection{Rosenbrock function} \label{sec:appl:2d}
\subsubsection{Problem definition}

The Rosenbrock function is a benchmark analytical (polynomial) function with two input variables used for optimization \citep{Rosenbrock1960}:
\begin{equation}
y = f_1(\vx) = 100\prt{x_2-x_1^2}^2 + (1-x_1)^2.
\end{equation}
Here, we model parameters $x_1 $ and $x_2$ by two independent p-boxes $X_1$ and $X_2$. 

\begin{figure}[ht!]
\centering
\subfigure[3D visualization of $f_1$\label{fig:four:c}]{
	\includegraphics[width=0.3\linewidth]{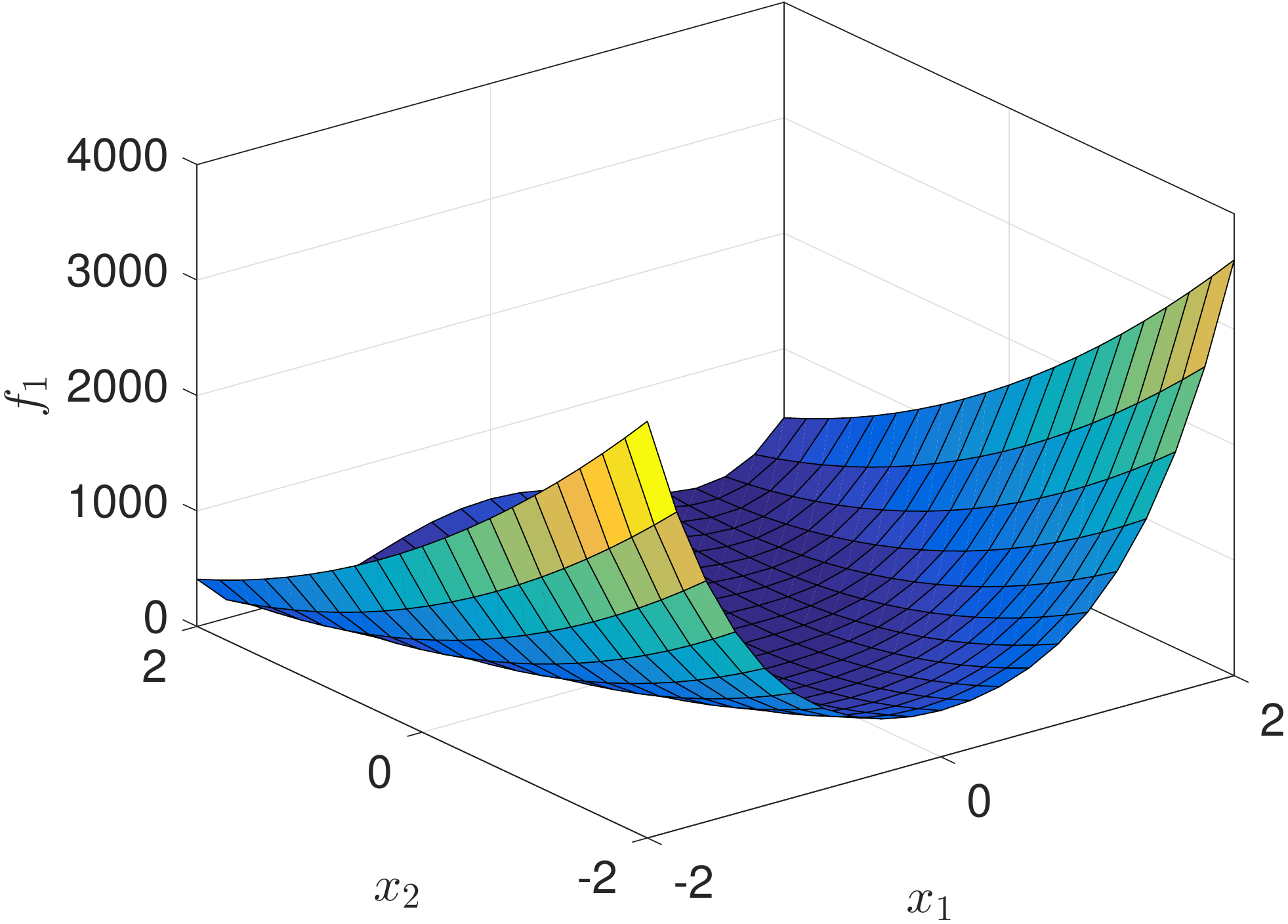}
}
\subfigure[Case \#1 -- p-box boundary curves for $X_i$ \label{fig:four:a}]{
	\includegraphics[width=0.3\linewidth]{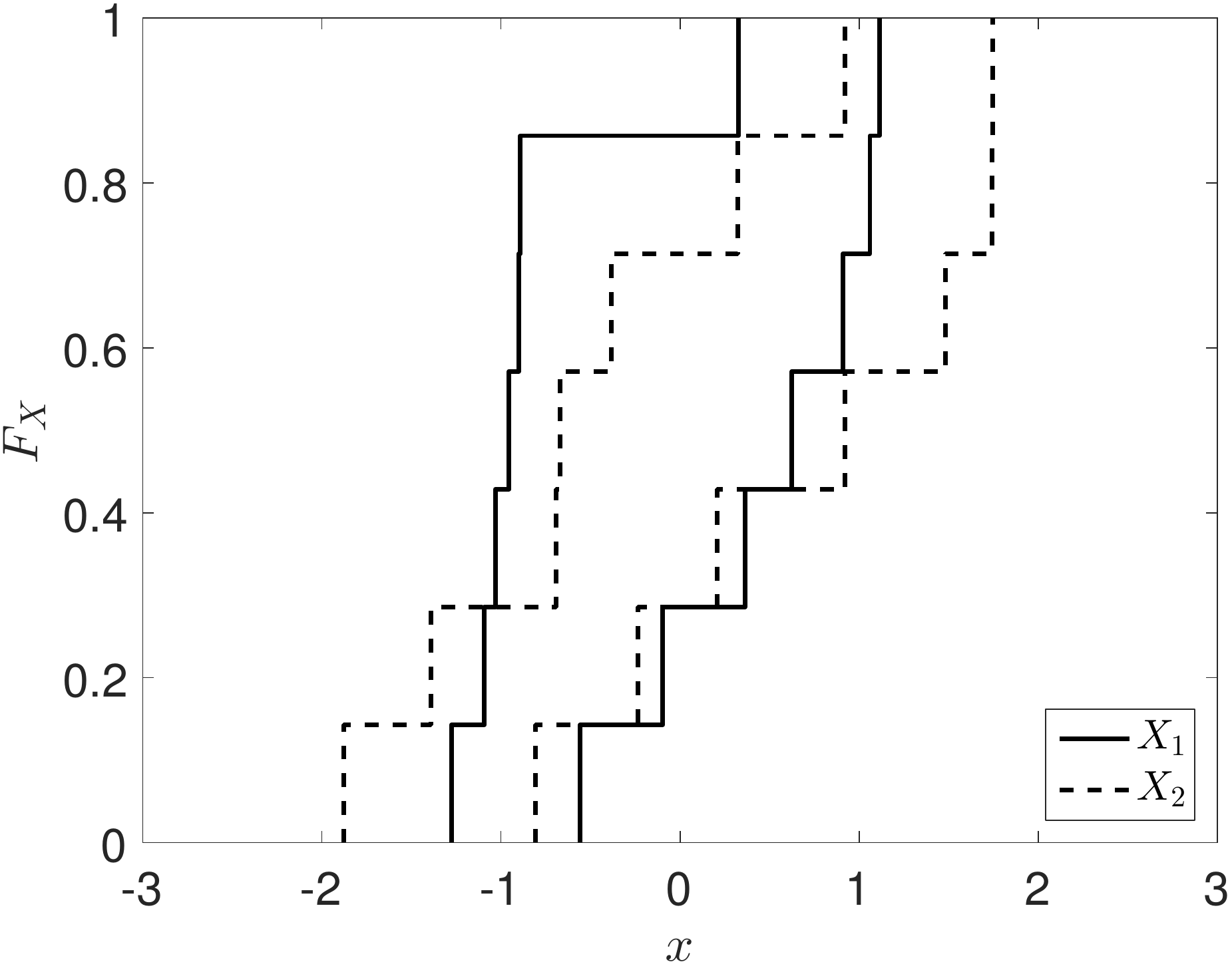}
}
\subfigure[Case \#2 -- p-box boundary curves for $X_i$ \label{fig:four:b}]{
	\includegraphics[width=0.3\linewidth]{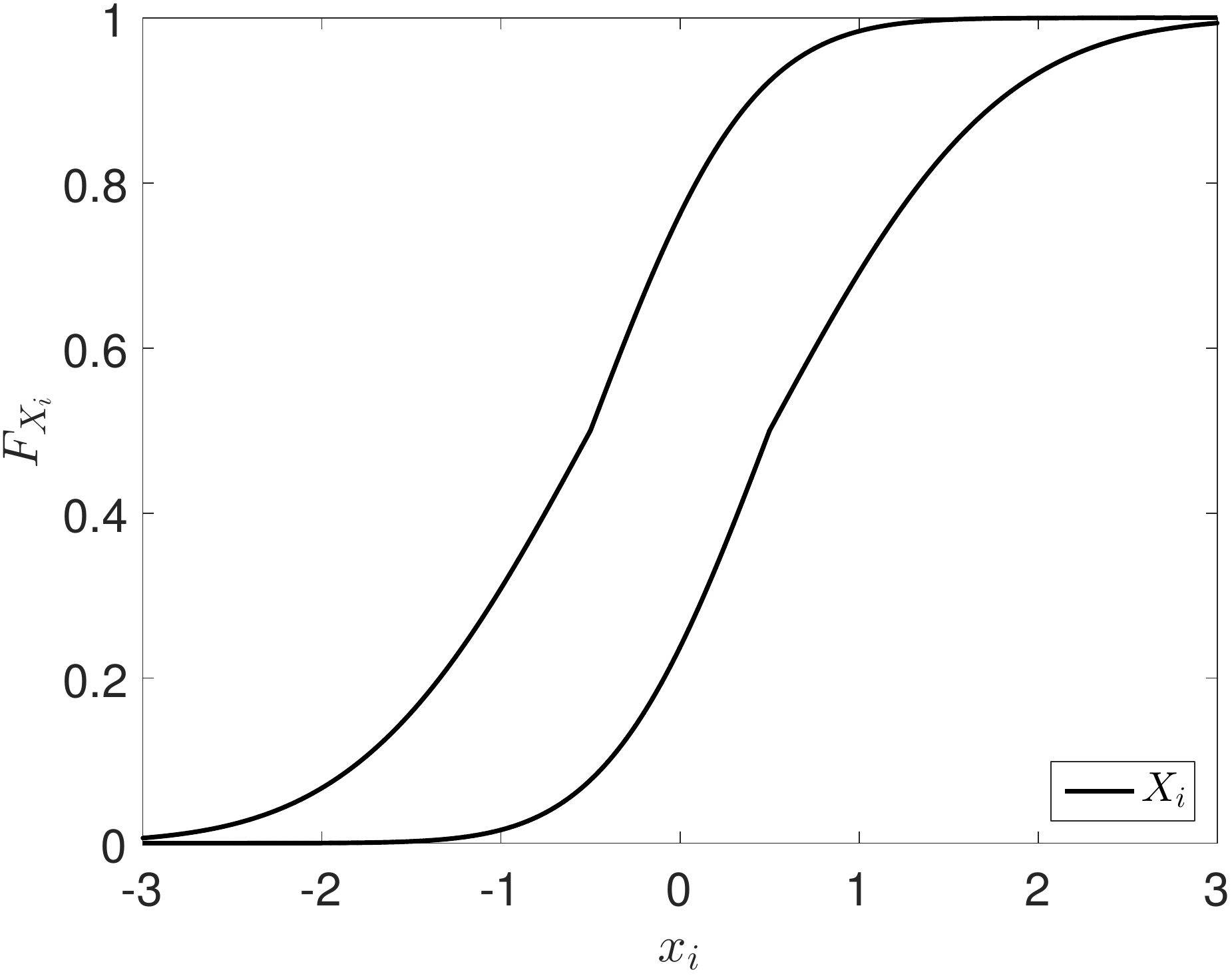}
}
\caption{Rosenbrock function -- Problem setup \label{fig:four}}
\end{figure}

Figure~\ref{fig:four:c} shows the response surface of $f_1$ as a function of $x_1$ and $x_2$. The response is non-monotone around the origin of the input space and has a global minimum of $y=0$ at $\acc{x_1=1,\,x_2=1}$. Thus, the QoI $Y$ is bounded on one side to the domain $\mathcal{D}_Y = [0,\infty]$. 
 
\paragraph{Case \#1}
The p-boxes based on the opinions of seven experts are illustrated in Figure~\ref{fig:four:a}. The solid lines mark the p-box for $X_1$, whereas the dashed lines mark the p-box for $X_2$. It can be seen that the credibility of the experts is uniform due to the constant vertical step size between the vertical plateaus of the boundary CDFs, and that the p-boxes are bounded on both sides.

\paragraph{Case \#2}
The bounds of the two input variables $X_i$ are defined as follows (Case \#2(b)):
$$\lF_{X_i}(x_i) = \min_{\mu\in[\underline{\mu},\overline{\mu}],\sigma\in[\underline{\sigma},\overline{\sigma}]}F_{\mathcal{N}}\prt{x_i\middle|\mu,\sigma},$$
$$\uF_{X_i}(x_i) = \max_{\mu\in[\underline{\mu},\overline{\mu}],\sigma\in[\underline{\sigma},\overline{\sigma}]}F_{\mathcal{N}}\prt{x_i\middle|\mu,\sigma},$$
where $F_{\mathcal{N}}(x|\mu,\sigma)$ is a Gaussian CDF with mean value $\mu$ and  standard deviation $\sigma$. In order to generate a p-box, the distribution parameters are given in intervals: $\mu\in\bra{\underline{\mu},\overline{\mu}}=\bra{-0.5,0.5}$ and $\sigma\in\bra{\underline{\sigma},\overline{\sigma}}=\bra{0.7, 1.0}$.
Figure~\ref{fig:four:b} shows the boundary curves of the p-boxes for $X_i$, $i=1,2$.

\subsubsection{Analysis}
%
The two-level meta-modelling approach has been implemented taking advantage of the \textsc{Matlab}-based uncertainty quantification framework \textsc{UQLab} \citep{MarelliICVRAM2014,UQdoc_09_104}. The sparse PCE meta-models were calibrated with Latin-hypercube experimental designs \citep{McKay1979} and appropriate sets of orthonormal polynomials (Legendre polynomials for bounded variables and Hermite polynomials for unbounded variables). The set of polynomials is determined by a degree-adaptive LAR of which the maximum total polynomial degree is set to $30$ and the parameter for the hyperbolic truncation set $\cA^{M,p}_q$ is set to $q=0.75$. The number of samples in the experimental design is denoted by $N_1$ and $N_2$ for constructing the meta-models of $\cm$ and $\acc{\lcm,\ucm}$, respectively. In order to achieve a statistical significance, the uncertainty propagation analyses are replicated 50 times with different LHS experimental designs.

For Case \#1, the auxiliary input variables are defined as in Eq.~(\ref{eq:auxU}), whereas for Case \#2, the auxiliary input variables are defined as $F_{\widetilde{X}_i} = F_{\mathcal{N}}(x|0,1)$. Note that for Case \#2, both levels of meta-models are trained in the same auxiliary domain, \ie $\mathcal{D}_{\widetilde{\vX}}$. 

The performance of the proposed two-level meta-modelling approach is measured in terms of the goodness-of-fit of the meta-models in the two levels of the approach. The accuracy of the meta-models is measured by the \emph{relative generalization error}, which is defined as follows:
\begin{equation}
err_{\text{gen}}\bra{Y} = \frac{\mathbb{E}\bra{\prt{Y-Y^{(P)}}^2}}{\rm{Var}\bra{Y}},
\end{equation}
where $Y$ is the true response variable and $Y^{(P)}$ is its meta-model response value. An estimate of the relative generalization error is computed by a large sample set $\acc{\ve{x}_i, \ i=1,\ldots,10^6}$, called validation set, as follows:
\begin{equation}
\widehat{err}_{\text{gen}}\bra{Y} \eqdef \frac{\sum_i\prt{{y}_i - {y}_i^{(P)}  }^2}{\sum_i\prt{{y}_i - \mu_{{Y}}}^2},
\end{equation}
where $y_i$ is the exact value of response of the computational model, $y_i^{(P)}$ is the prediction value based on the meta-model, and $\mu_Y$ is the estimated mean value of the response variable $Y$.

A second measure for the accuracy of the proposed two-level approach is the comparison of the true and the meta-modelled p-boxes for the QoI $Y$. In particular, the graphical comparison of the boundary CDFs $\underline{Y}$ and $\overline{Y}$ is made in the typical variable-CDF plot and shows the quality of the proposed approximations. {To numerically quantify the convergence of the results, the following p-box area is defined:}
{
\begin{equation}
A_Y = \int_{-\infty}^{\infty} \prt{\overline{F}_Y(y) - \underline{F}_Y(y)} \, \text{d}y.
\end{equation}
}
{Then, the approximated p-box area is compared to the true p-box area. Additionally, the Kolmogorov-Smirnov distance of the bounds and their approximations is computed and combined to:}
{\begin{equation}
D = \left|D_{\underline{Y}}\right| + \left|D_{\overline{Y}}\right|,
\end{equation}}
{where $D_{\underline{Y}}$ and $D_{\overline{Y}}$ are the KS distances between the reference solution and the approximate solutions for $\underline{Y}$ and $\overline{Y}$, respectively, defined as:}
{\begin{equation}
D_{\underline{Y}} = \max_{y\in\cD_Y} \left| \underline{F}_{Y,ref}(y)-\underline{F}_Y^{(\text{P})}(y) \right|, \qquad D_{\overline{Y}} = \max_{y\in\cD_Y} \left|\overline{F}_{Y,ref}(y)-\overline{F}_Y^{(\text{P})}(y) \right|,
\end{equation}}
{where $\underline{F}_{Y,ref}$ (resp. $\overline{F}_{Y,ref}$) is the CDF of the reference solution and $\underline{F}_Y^{(\text{P})}$ (resp. $\overline{F}_Y^{(\text{P})}$) is its approximation.}

\subsubsection{Results}
\paragraph{Case \#1}
The polynomial form of the Rosenbrock function can be modelled exactly with the polynomial-based PCE models. Hence, the first-level meta-model does not introduce any approximation in this application example. 

\begin{figure}[ht!]
\centering
\subfigure[Response $\underline{Y}$ (\ie based on $\ucm$) for $N_1 = 30$ \label{fig:four:results1:b}]{
	\includegraphics[width=0.3\linewidth]{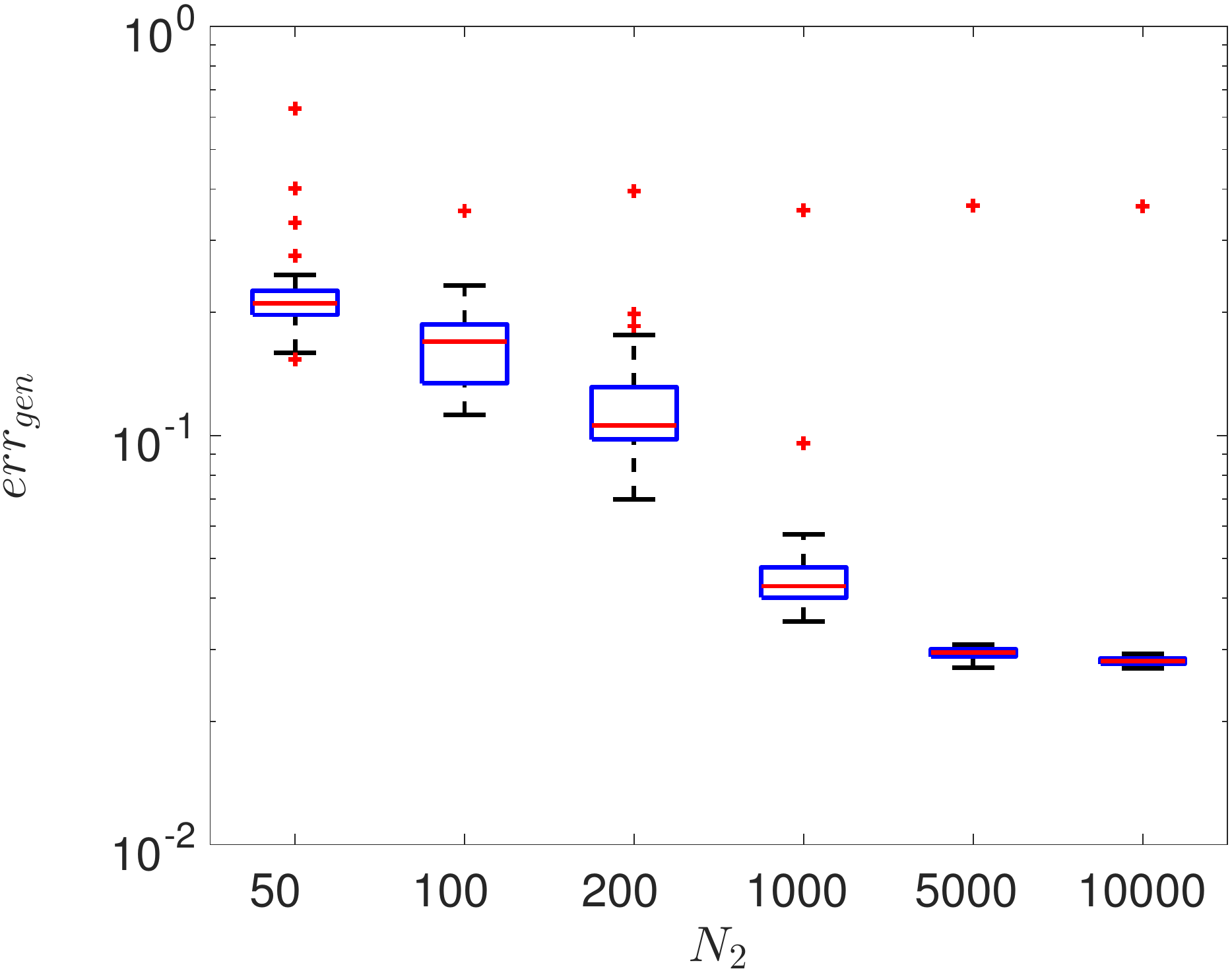}
}
\hspace{1cm}
\subfigure[Response $\overline{Y}$ (\ie based on $\lcm$) for $N_1 = 30$ \label{fig:four:results1:c}]{
	\includegraphics[width=0.3\linewidth]{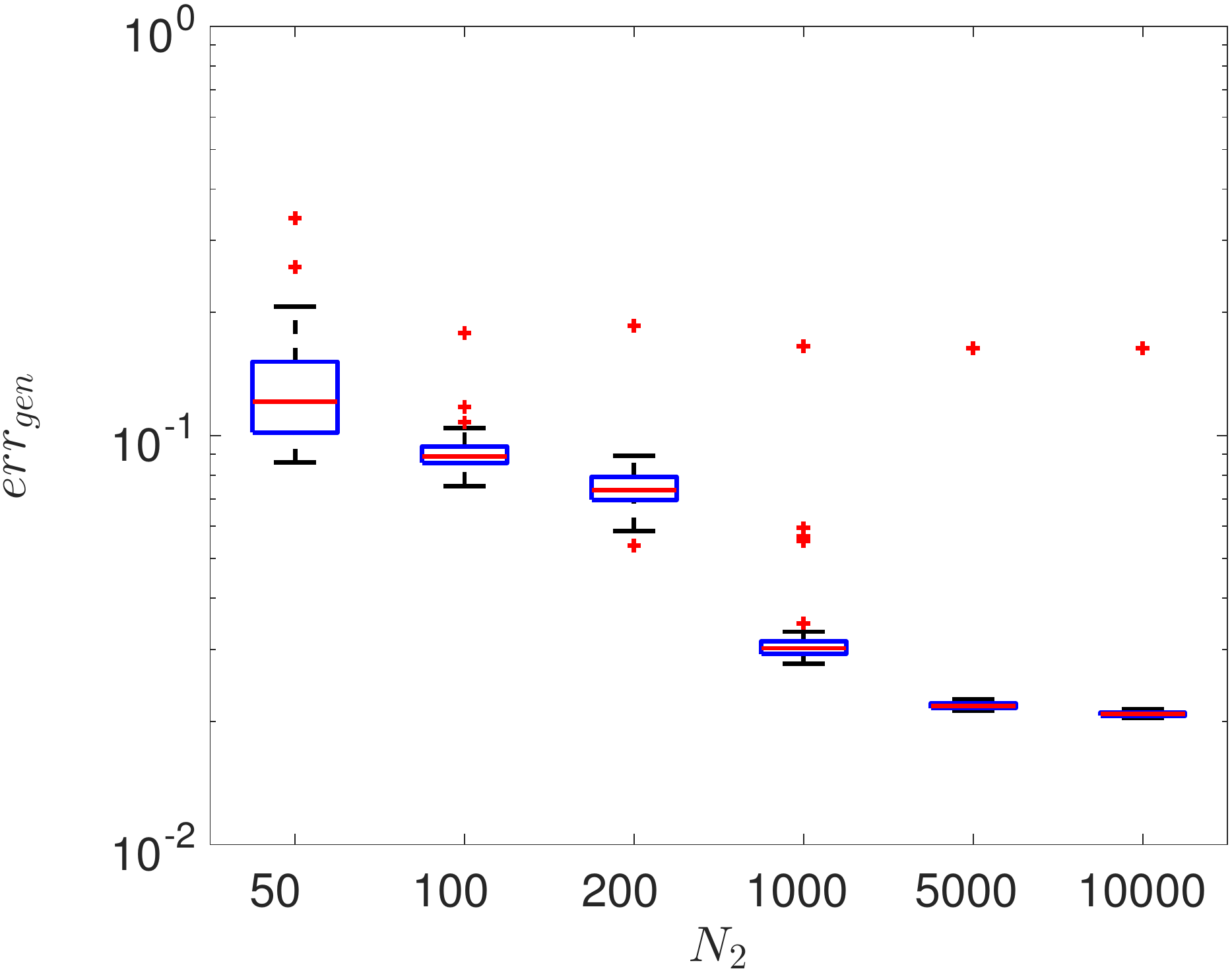}
}
\caption{Rosenbrock function -- Case \#1 -- relative generalization error as a function of $N_1,\,N_2$ and based on $50$ replications of the same analysis \label{fig:four:results1}}
\end{figure}

The reference solution for the second-level models $\lcm$ and $\ucm$ is obtained by propagating all possible combinations of input intervals, \ie $n_{E,X_1} \cdot n_{E,X_2} = 7\cdot 7 = 49$ intervals. This results in stairs-shaped boundary curves for the response p-box. These shapes are difficult to model with polynomials, which generally define smooth functions. Hence, the accuracy of the second-level meta-models is low, as see in Figure~\ref{fig:four:results1}, which summarizes the relative generalization error for the two second-level meta-models: $\ucm^{(P)}$ and $\lcm^{(P)}$. In fact, it requires  $N_2=1,000$ samples to achieve an accuracy of $\widehat{err}_{\text{gen}}\approx 0.05$ due to the shape of the input p-boxes. In other words, the continuous input vector $\vC$ is mapped onto $49$ values in the response variables $\underline{Y}$ and $\overline{Y}$.

\begin{figure}[ht!]
\centering
\subfigure[ $\cm$ \label{fig:rosen:response:1}]{
\includegraphics[width = 0.3\linewidth]{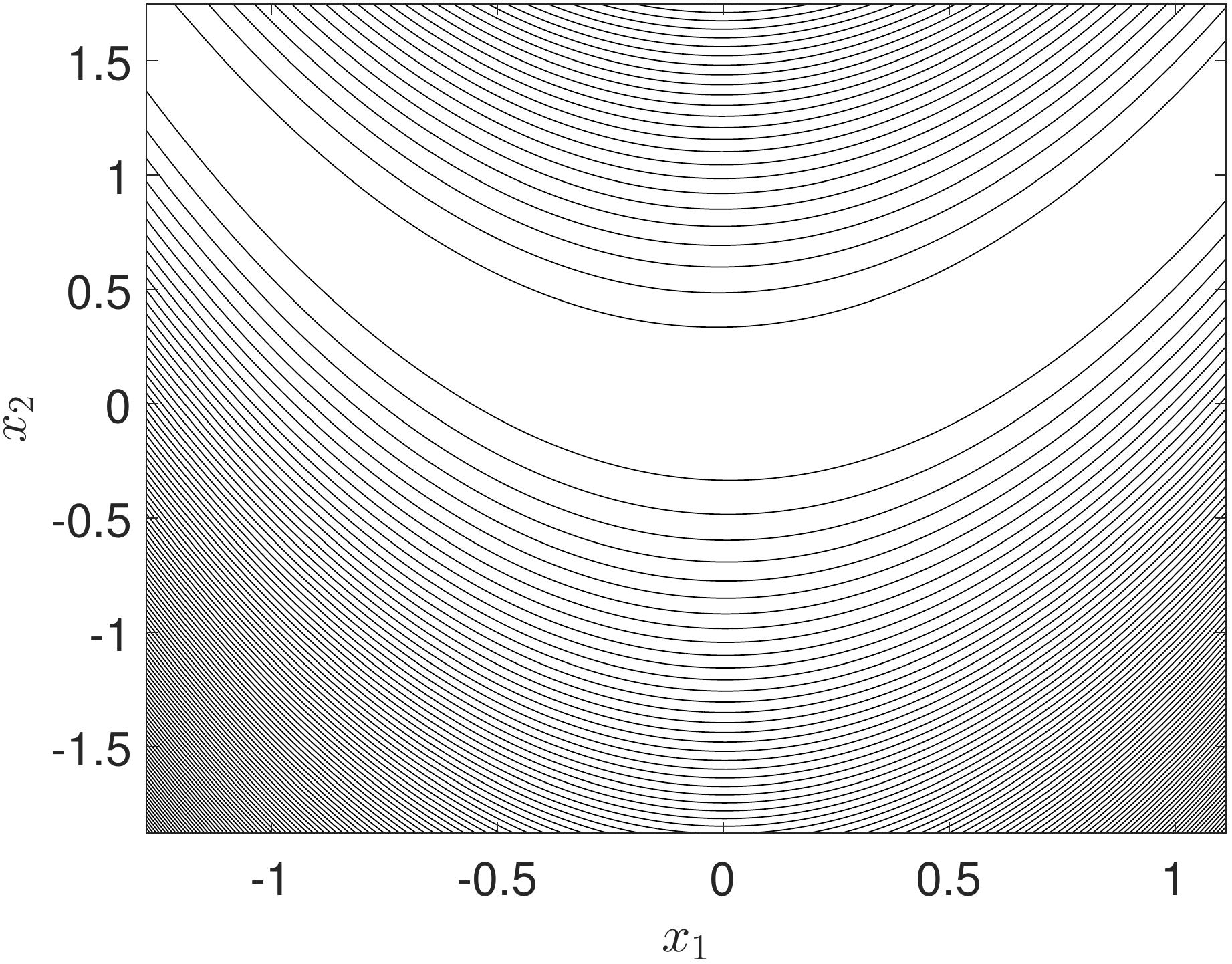}
}
\subfigure[ $\ucm$ \label{fig:rosen:response:2}]{
\includegraphics[width = 0.3\linewidth]{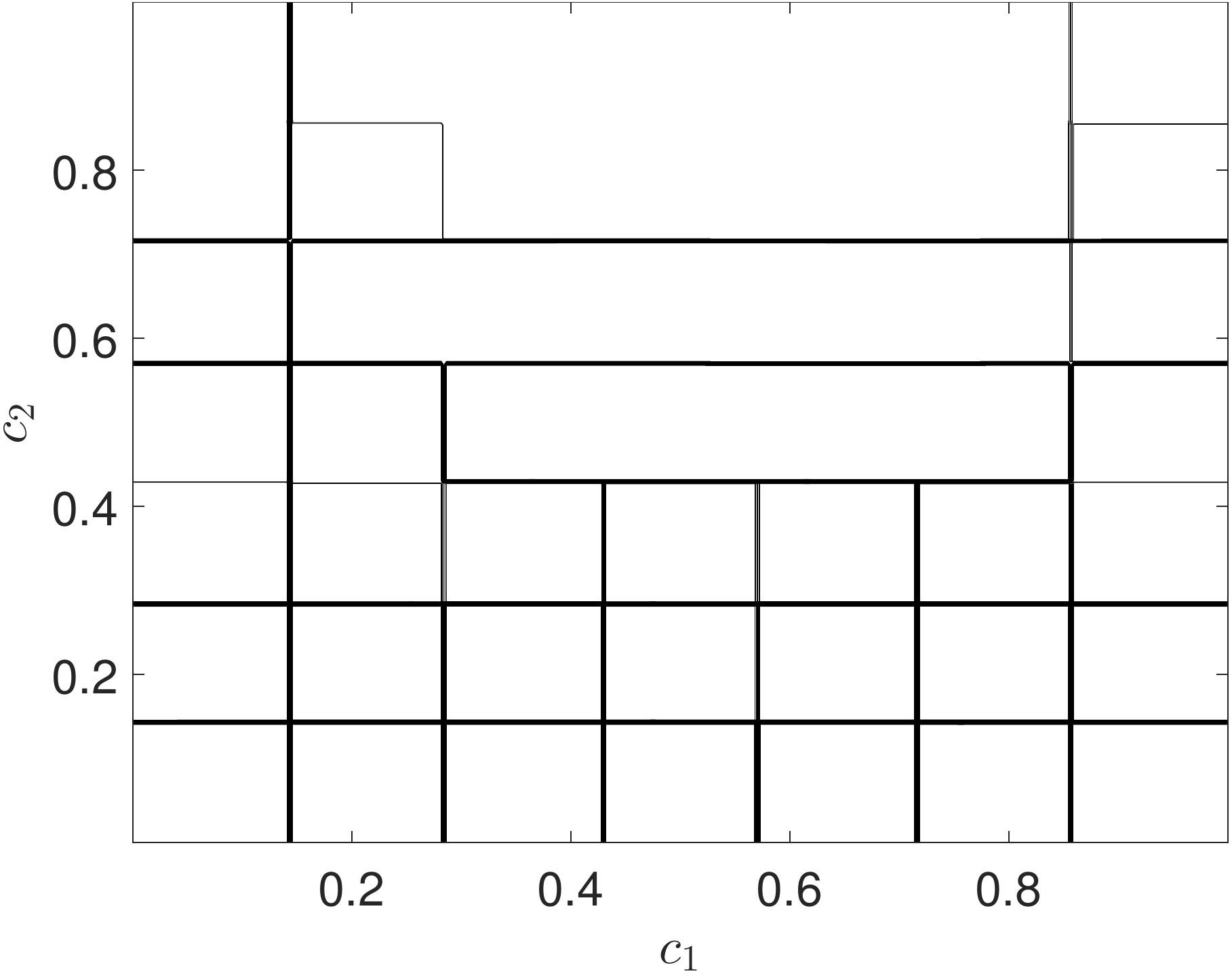}
}
\subfigure[ $\lcm$ \label{fig:rosen:response:3}]{
\includegraphics[width = 0.3\linewidth]{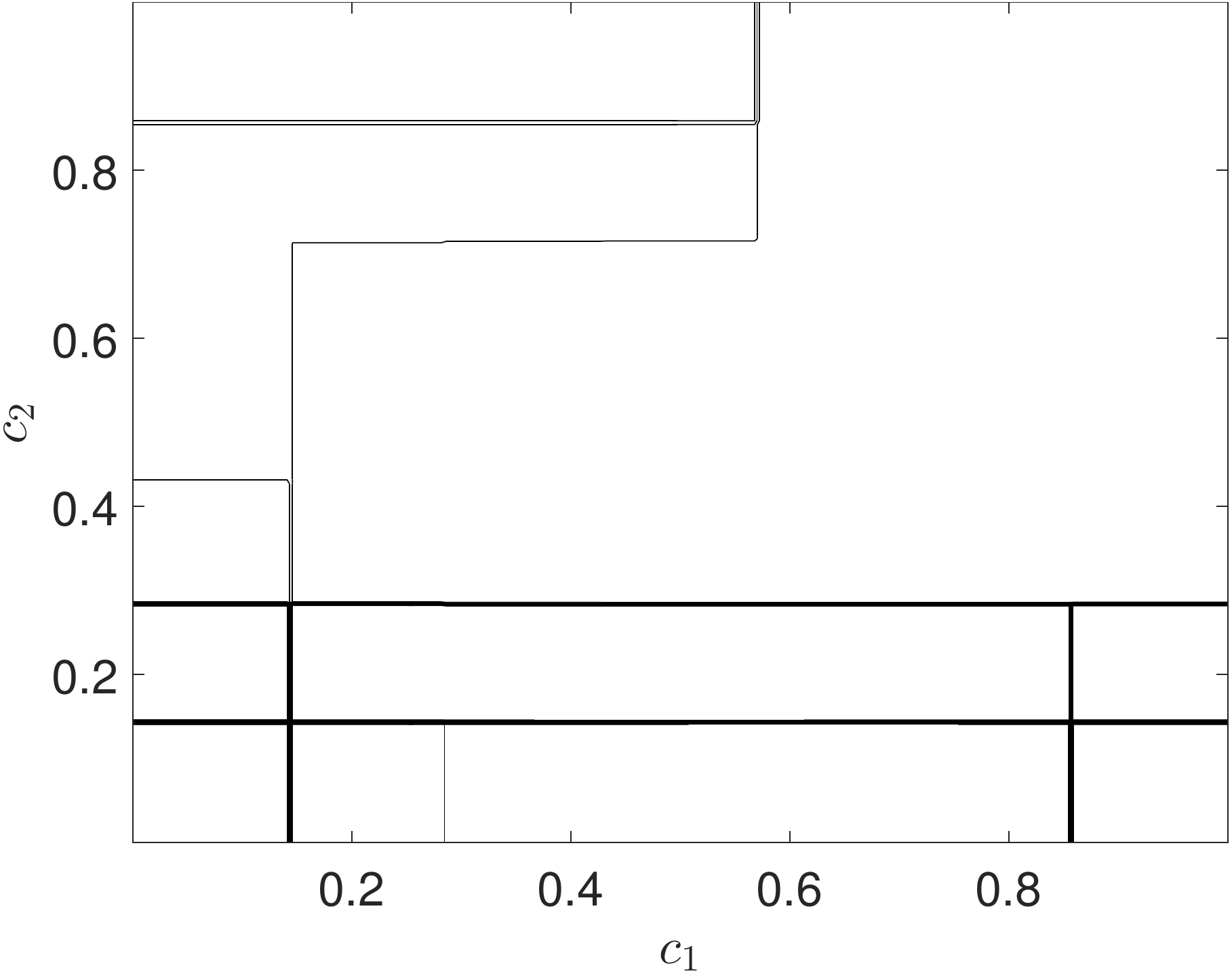}
}
\caption{Rosenbrock function -- Case \#1 -- shape of the response surfaces \label{fig:rosen:response}}
\end{figure}

The explanation for this convergence behaviour is illustrated in Figure~\ref{fig:rosen:response}, which shows the response surface of the computational models as a function of the input random variables. The computational model $\cm$, which is the basis for the first-level meta-model, shows smooth contour lines (see Figure~\ref{fig:rosen:response:1}), whereas $\lcm$ and $\ucm$ show plateau-shaped response surfaces (see Figure~\ref{fig:rosen:response:2} and \ref{fig:rosen:response:3}) with constant response values within each plateau. A total of $7\times 7 = 49$ distinct plateaus can be identified corresponding to the $49$~possible combinations of input intervals defined by the expert opinions. Instead of training second-level meta-models, the original (small) number of input intervals, \ie the intervals given by the experts, may be propagated directly to estimate the bounds of the response p-box. In other words, the slicing algorithm may be applied of the second-level of the uncertainty analysis, due to the low number of response intervals, \ie $|\cK|=49$.

\begin{figure}[ht!]
\centering
	\includegraphics[width=0.4\linewidth]{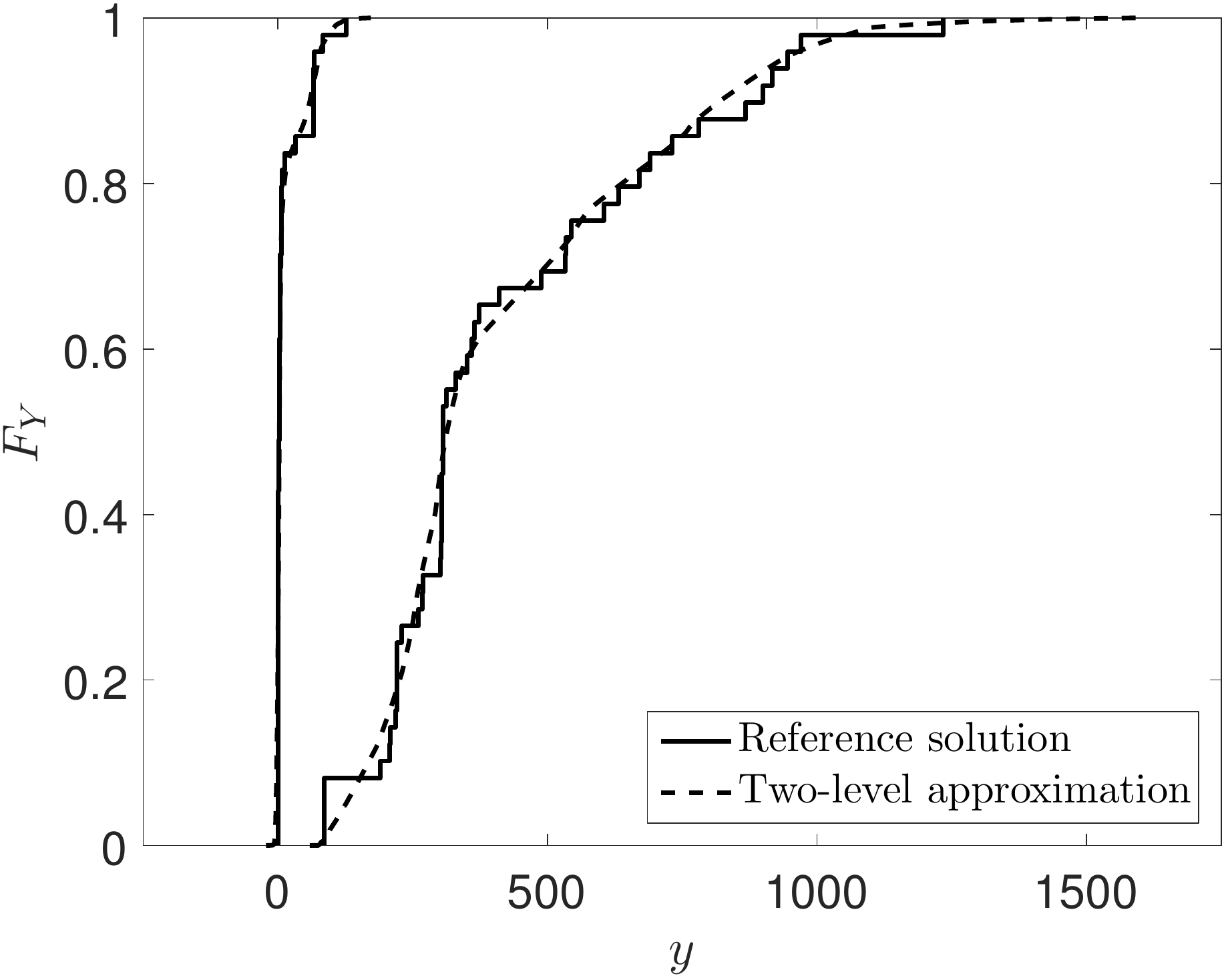}
\caption{Rosenbrock function -- Case \#1 -- Response p-boxes $Y\in\bra{\underline{Y}, \overline{Y}}$: reference solution ($n=7\cdot 7=49$) versus two-level approach ($N_1 = 50$, $N_2 = 200$) \label{fig:four:results:1:a}}
\end{figure}

When comparing the boundary curves of the response p-box, however, results show that the proposed approach provides accurate (although smooth) approximations of the response p-box boundaries.
The exact boundary curves of the p-box of $Y$ are shown in Figure~\ref{fig:four:results:1:a}, where the distinct values are clearly visible in the stairs-shaped CDF curves. For comparison, the boundary curves of the two-level approach are presented for a single run with $N_1 = 50$ and $N_2 = 200$. For this run, the relative generalization error are $\widehat{err}_{\text{gen}}\bra{\overline{Y}} = 1.14\cdot 10^{-1}$ and $\widehat{err}_{\text{gen}}\bra{\underline{Y}} = 8.76\cdot 10^{-2}$. The PCE-based CDFs are not capable of reproducing the stairs-like functions but still follow the exact boundary curves sufficiently accurately for visual comparison. This explains the large relative generalization errors in the second-level meta-models.

\begin{figure}[ht!]
\centering
\subfigure[\label{fig:four:results:ai}{P-box area}]{
\includegraphics[width=0.3\linewidth]{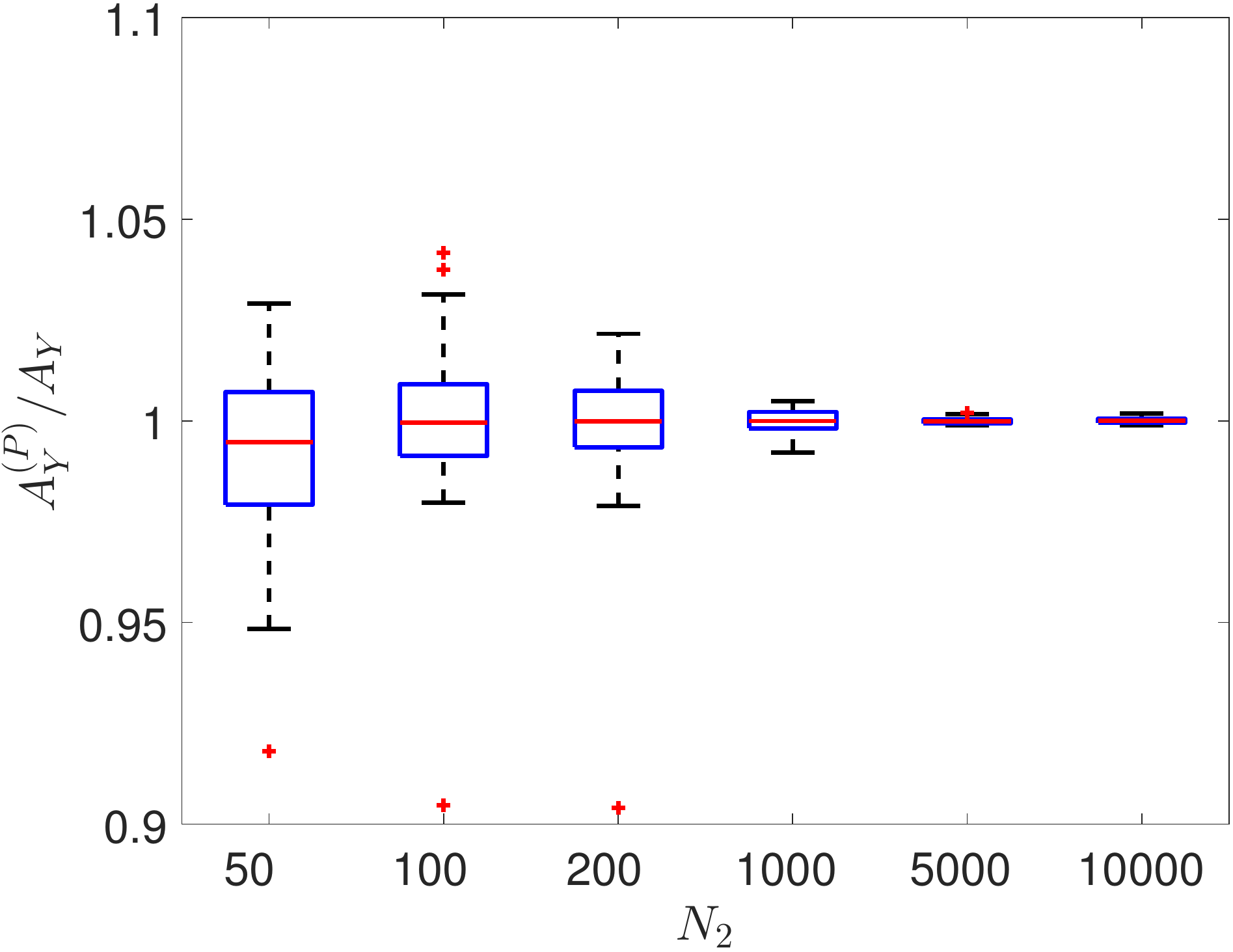}
}
\subfigure[\label{fig:four:results:ks}{KS distance}]{
\includegraphics[width=0.3\linewidth]{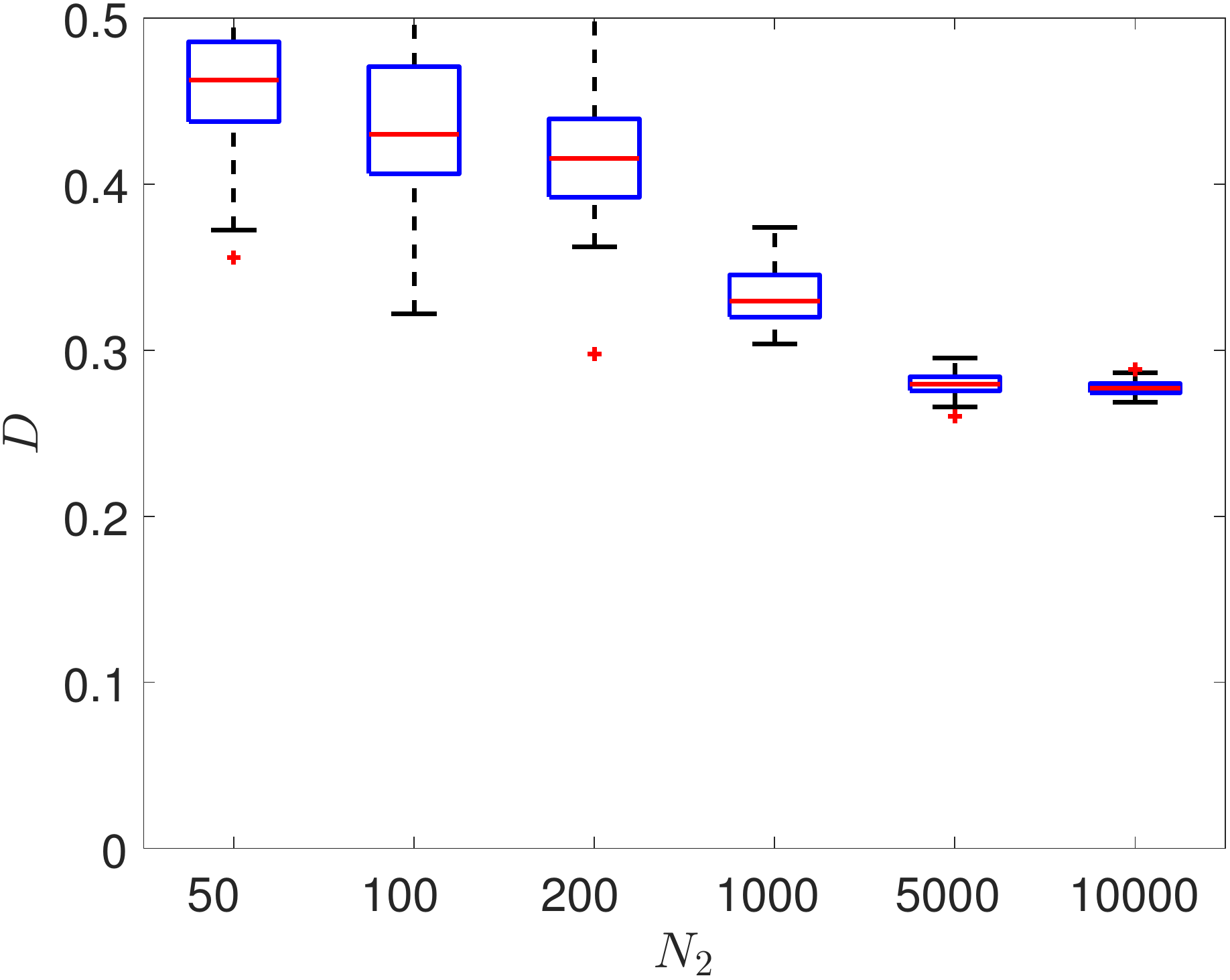}
}
\caption{{Rosenbrock function -- Case \#1 -- Convergence of response p-box as a function of $N_2$ and based on 50 replications of the same analysis}}
\end{figure}

{The quantitative values of $A_Y^{(\text{P})}/A_Y$ are shown in Figure~\ref{fig:four:results:ai} together with the KS distances in Figure~\ref{fig:four:results:ks}. The two figures nicely illustrate that the increase of samples in the experimental design improves the accuracy of the two-level meta-modelling algorithm. As the number of samples increases, the response p-box area is modelled more and more accurately. The KS distance measure, however, does not converge as rapidly. The reason for the low convergence rate lies in the shape of the response p-box seen in Figure~\ref{fig:four:results:1:a}, in particular the shape of $\overline{F}_Y$. }

\begin{figure}[ht!]
\centering
\subfigure[Response $\underline{Y}$ (\ie based on $\ucm$) \label{fig:four:results2:b} with $N_1 = 100$]{
	\includegraphics[width=0.3\linewidth]{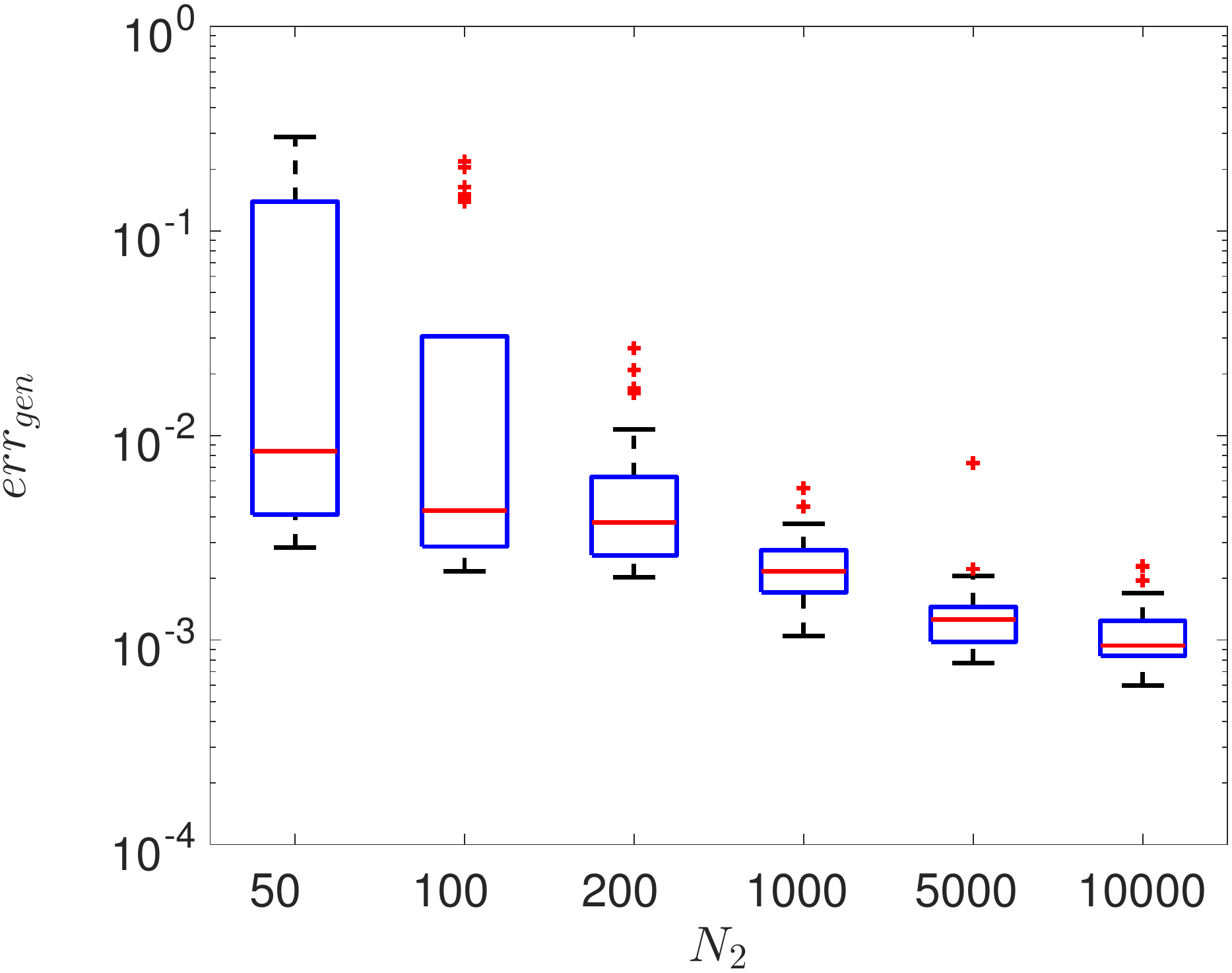}
}\hspace{1cm}
\subfigure[Response $\overline{Y}$ (\ie based on $\lcm$) \label{fig:four:results2:c} with $N_1 = 100$]{
	\includegraphics[width=0.3\linewidth]{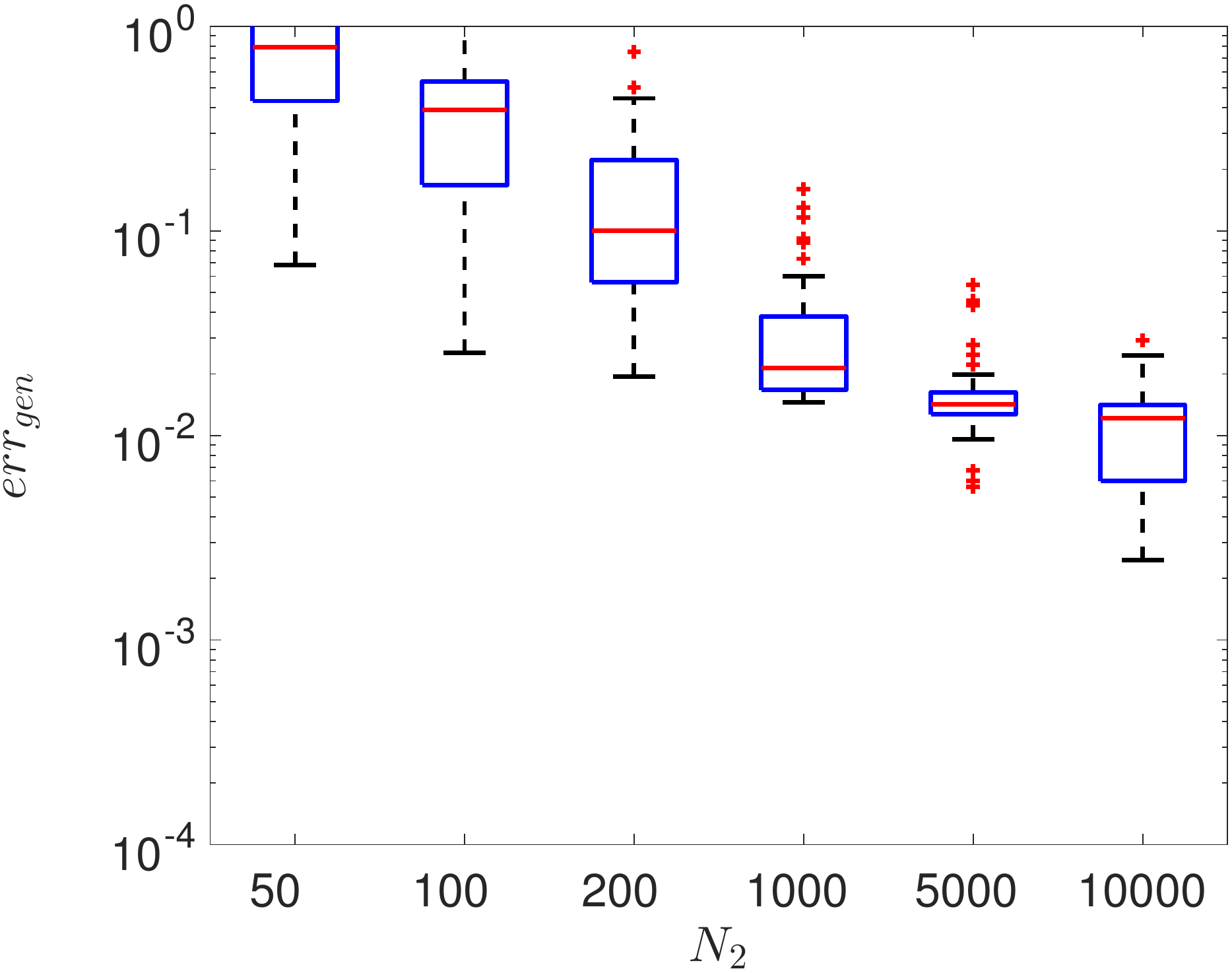}
}
\caption{Rosenbrock function -- Case \#2 -- relative generalization error as a function of $N_1,\,N_2$ and based on $50$ replications of the same analysis \label{fig:four:results2}}
\end{figure}

\paragraph{Case \#2}
%
Analogously to Case \#1, the performance of the second-level meta-models is shown in the two Figures~\ref{fig:four:results2:b} and \ref{fig:four:results2:c} for $\underline{Y}$ and $\overline{Y}$, respectively, using $N_1 = 100$ and {$N_2=\acc{50,\ldots, 10^4}$}. As expected, the relative generalization error becomes smaller with larger experimental design, \ie larger $N_2$. Interestingly however, the values are considerably different for $\underline{Y}$ and $\overline{Y}$, due to the shape of the response function $\ucm$ and $\lcm$, respectively.   

\begin{figure}[ht!]
\centering
\subfigure[$\cm$ \label{fig:four:d}]{
	\includegraphics[width=0.3\linewidth]{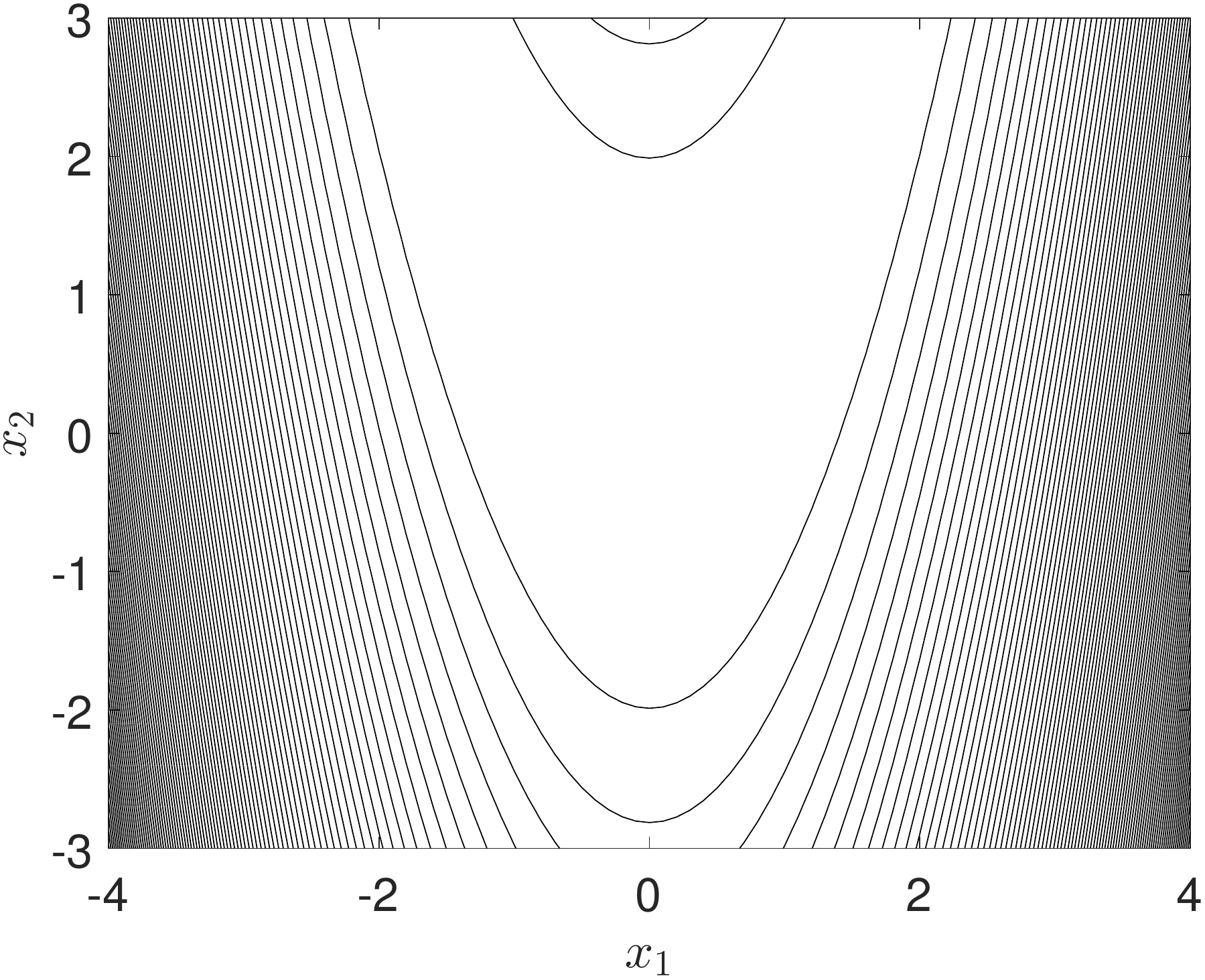}
}
\subfigure[$\ucm$ \label{fig:rosen:ucm}]{
	\includegraphics[width=0.3\linewidth]{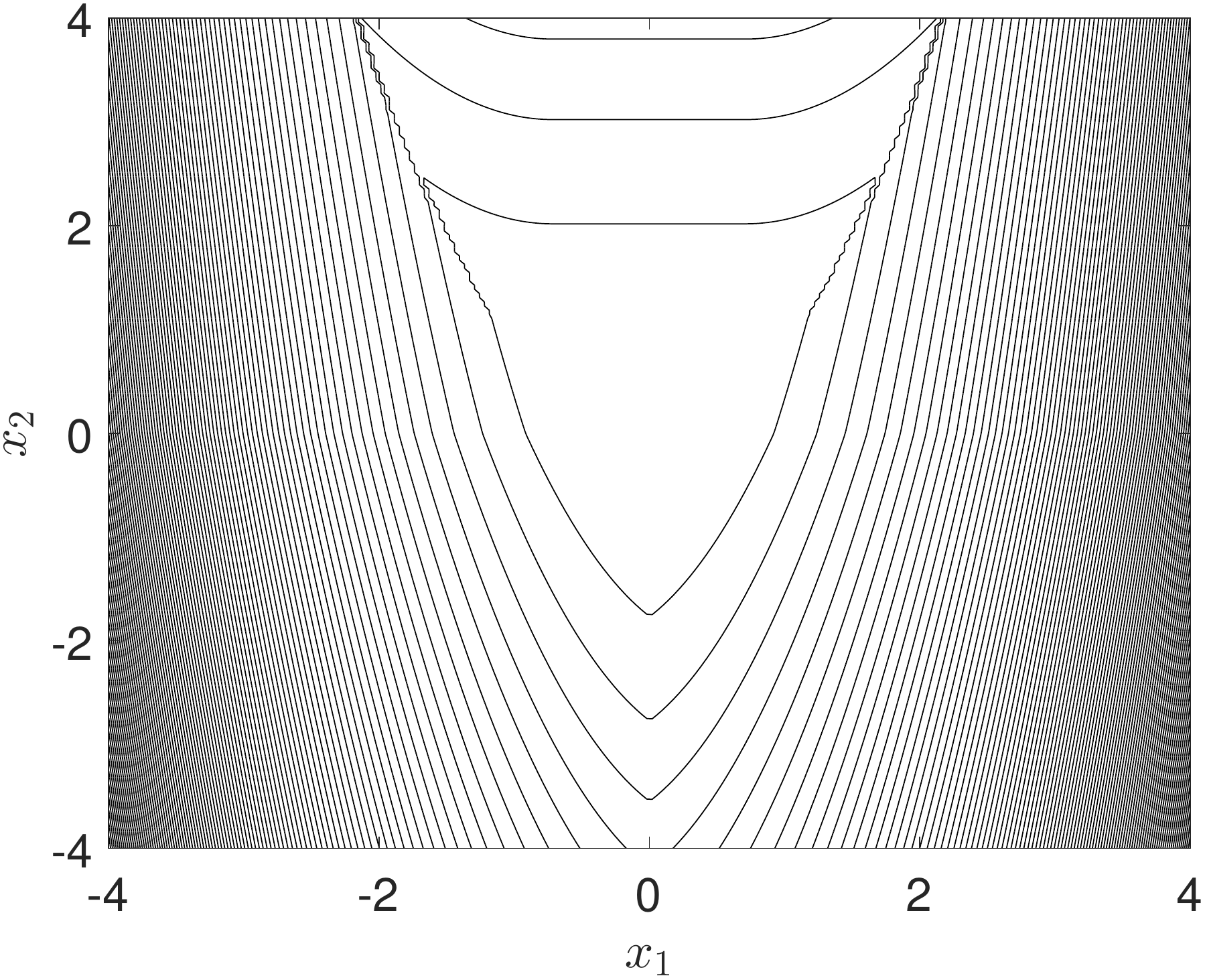}
}
\subfigure[$\lcm$ \label{fig:rosen:lcm}]{
	\includegraphics[width=0.3\linewidth]{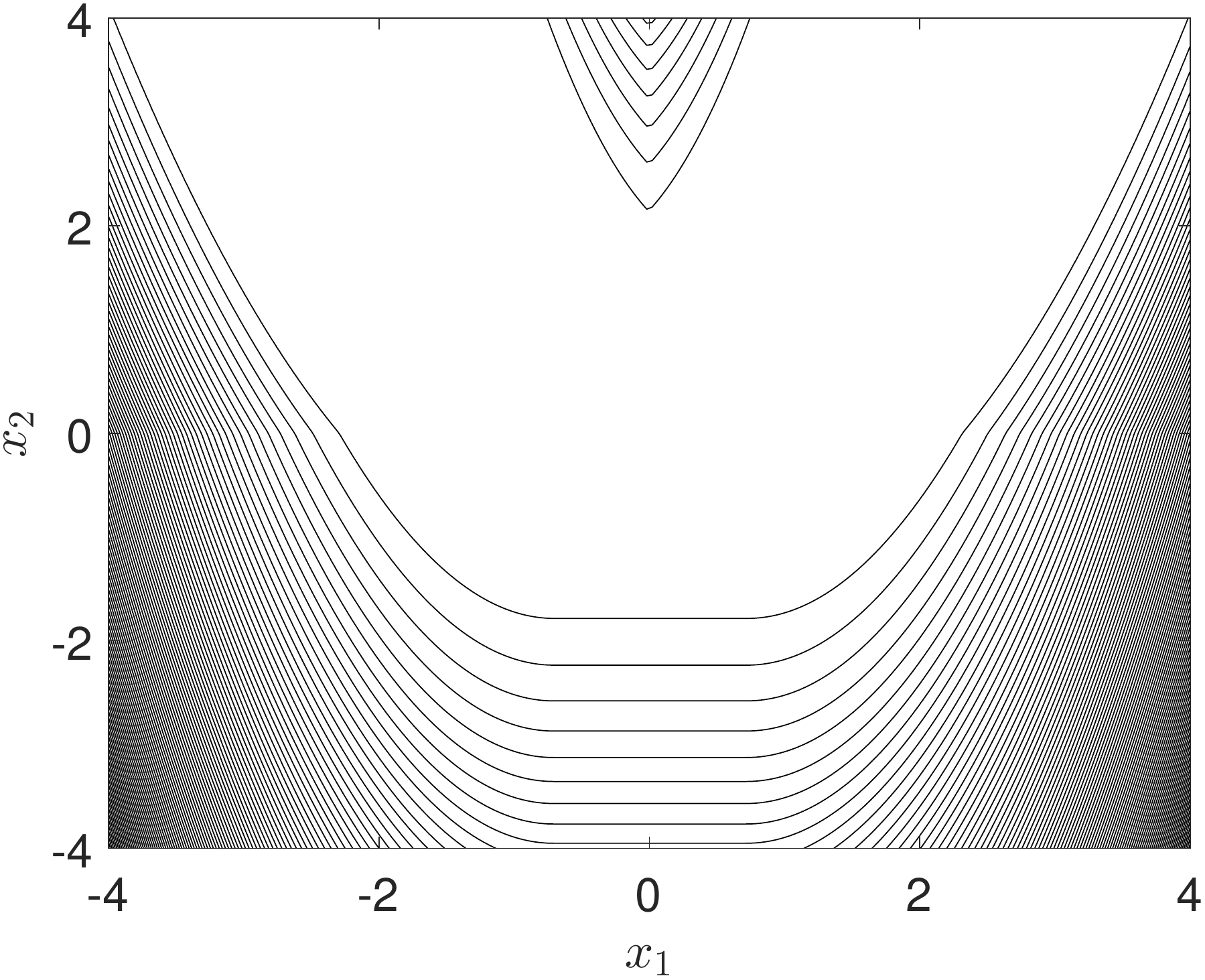}
}
\caption{Rosenbrock function -- shape of the response surfaces of $\lcm$ and $\ucm$ as a function of $\widetilde{\vX}$ \label{fig:rosen}}
\end{figure}

Figure~\ref{fig:rosen} shows the response surface for $\cm$, $\lcm$ and $\ucm$, respectively, as a function of the auxiliary input variables $\widetilde{\vX}$. Comparing the three sets of contour lines in Figure~\ref{fig:rosen}, the response surface of $\lcm$ has a large plateau around the origin and steep value increases around the plateau. This behaviour is difficult to model with polynomials, which results in a lower accuracy seen in Figure~\ref{fig:four:results2:c}. 

\begin{figure}[ht!]
\centering
\includegraphics[width=0.4\linewidth]{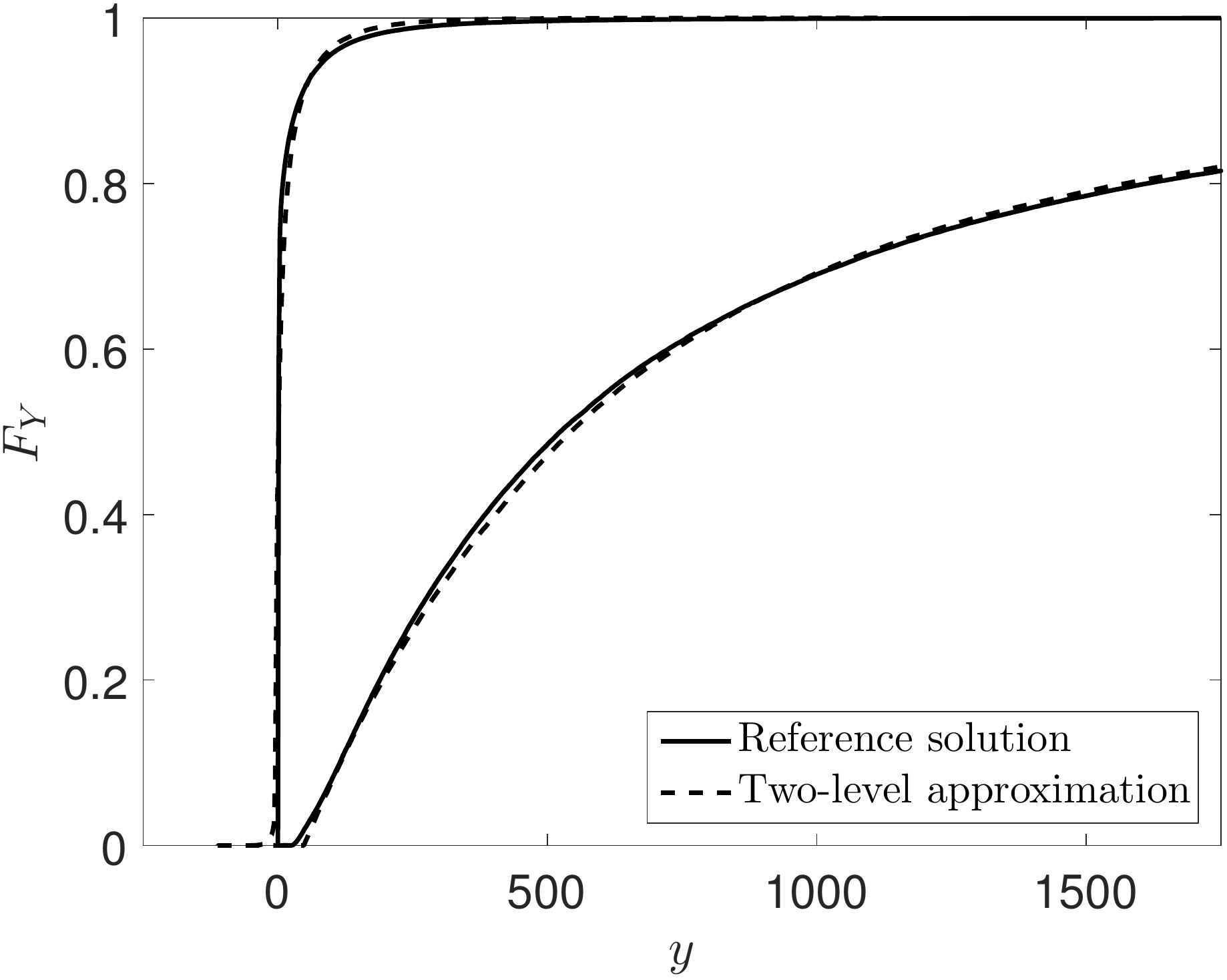}
\caption{Rosenbrock function -- Case \#2 -- Response p-boxes $Y\in\bra{\underline{Y}, \overline{Y}}$: reference solution ($n=10^6$) versus two-level approach ($N_1 = 100$, $N_2 = 200$) \label{fig:four:results:2}}
\end{figure}

Finally, the resulting boundary curves of the p-box of $Y$ are shown in Figure~\ref{fig:four:results:2}. The reference solution was computed with the exact computational model on the first level of the algorithm and $n=10^6$ samples on the second level. The effect of the identified plateau in $\lcm$ in Figure~\ref{fig:rosen:lcm} is that a large quantile of $\overline{Y}$ is close to $y=0$ (see Figure~\ref{fig:four:results:2}). Additionally, a single realization of those boundary curves is drawn in Figure~\ref{fig:four:results:2} for the case of $N_1 = 100$ and $N_2 = 200$. 
The corresponding relative generalization error for $\underline{Y}$ and $\overline{Y}$ are  $\widehat{err}_{\text{gen}}\bra{\overline{Y}} = 3.43\cdot 10^{-1}$ and $\widehat{err}_{\text{gen}}\bra{\underline{Y}} = 5.66\cdot 10^{-3}$, respectively. 
Despite the relatively large values of the relative generalization error of the second-level meta-model, the  bounds of the response p-box are modelled accurately. 

\begin{figure}[ht!]
\centering
\subfigure[\label{fig:rosen:results:case2:ai}{P-box area}]{
\includegraphics[width=0.3\linewidth]{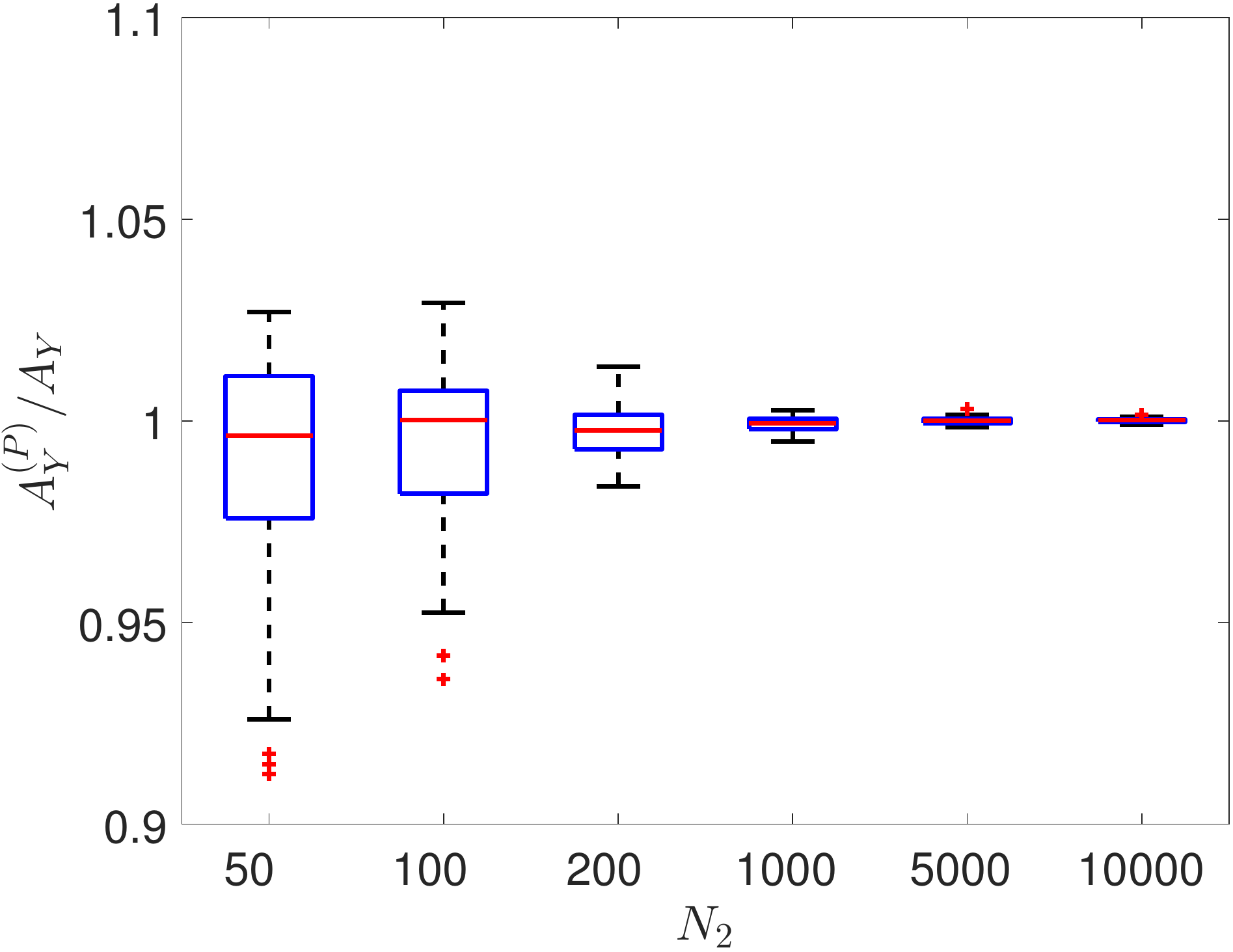}
}
\subfigure[\label{fig:rosen:results:case2:ks}{KS distance}]{
\includegraphics[width=0.3\linewidth]{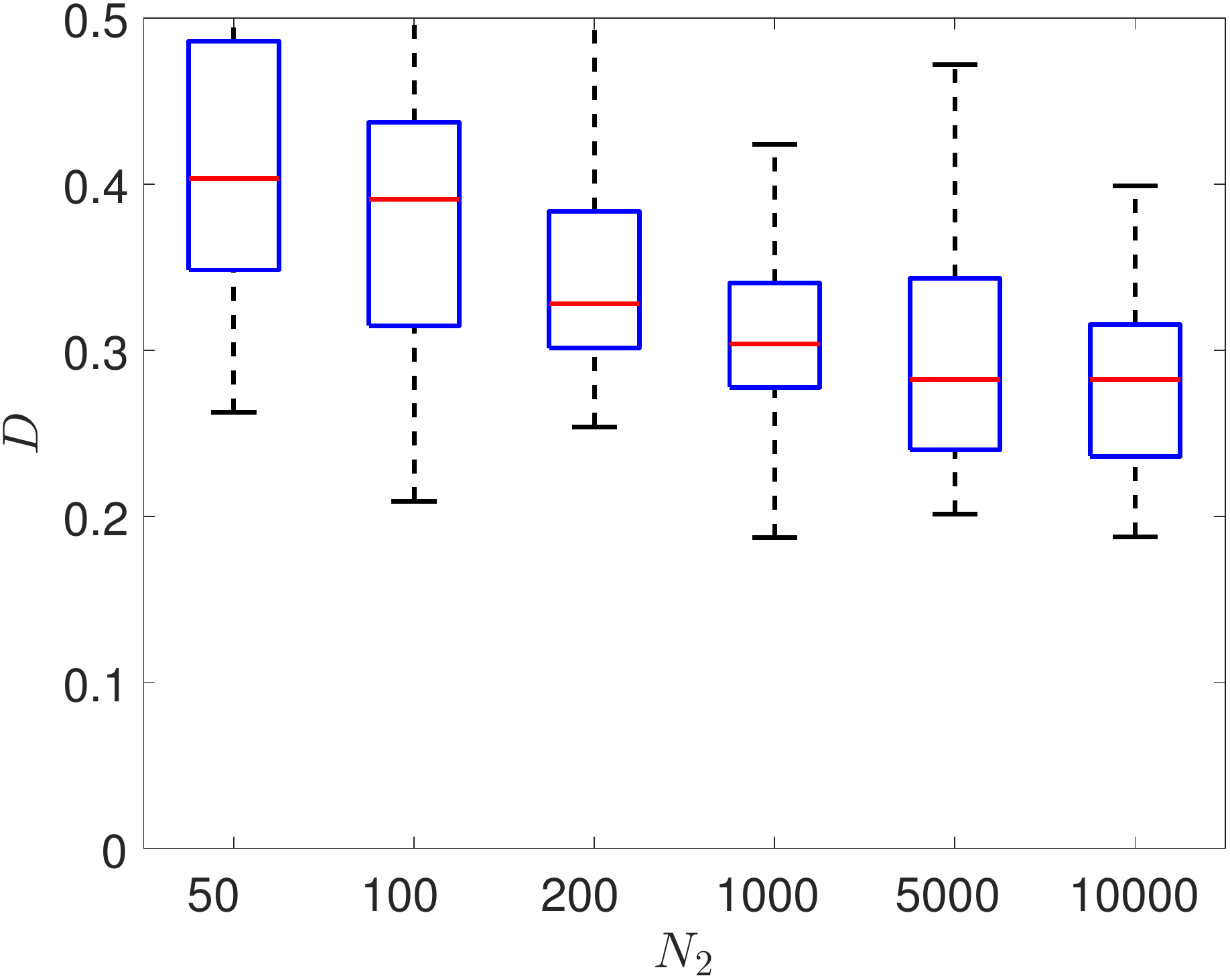}
}
\caption{{Rosenbrock function -- Case \#2 -- Convergence of response p-box as a function of $N_2$ and based on 50 replications of the same analysis}}
\end{figure}

{Figures~\ref{fig:rosen:results:case2:ai} and \ref{fig:rosen:results:case2:ks} illustrate the corresponding p-box area comparison and KS distance estimates, respectively. As in Case~\#1, they nicely show the convergence of the measures when increasing the number of samples in the experimental design of the second-level meta-model. Again, the p-box area converges nicely, whereas the KS distance converges more slowly, due to the shape of $\overline{F}_Y$ seen in Figure~\ref{fig:four:results:2}.}

\subsection{Two-degree-of-freedom damped oscillator} \label{sec:appl:nd}

\subsubsection{Problem definition}
\begin{figure}[ht!]
\centering
\includegraphics[width=0.6\linewidth]{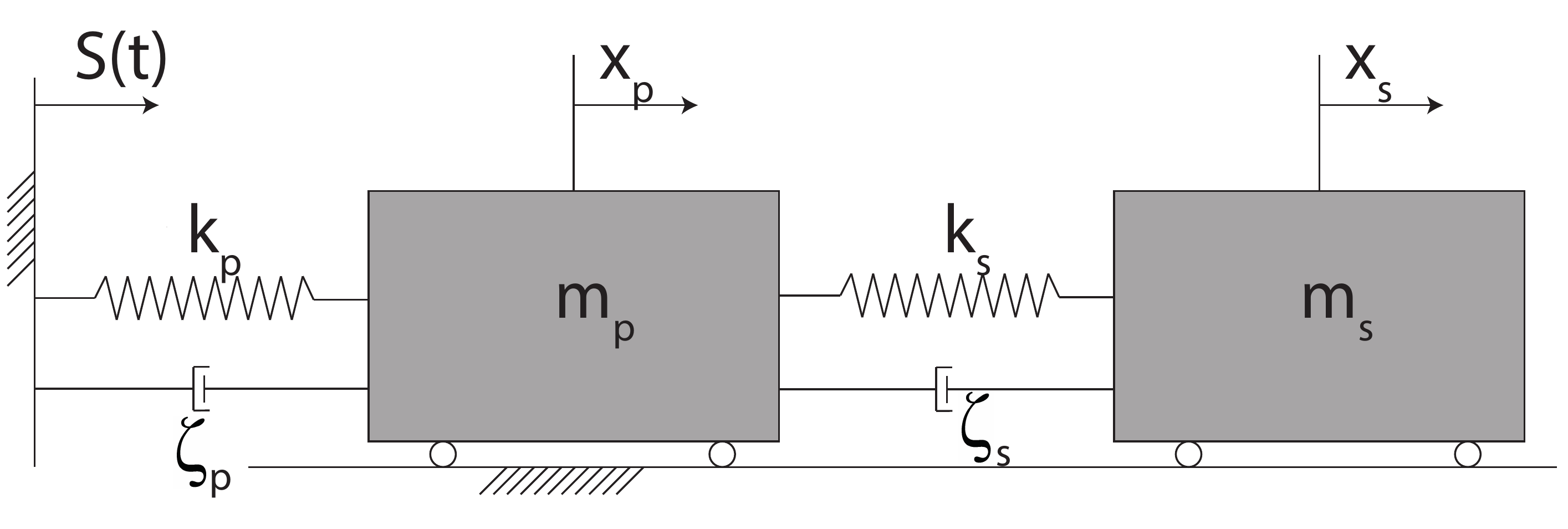}
\caption{\label{fig:damp:sketch} Sketch of the 2-dof damped oscillator}
\end{figure}

\begin{figure}[ht!]
\centering
	\includegraphics[width=0.4\linewidth]{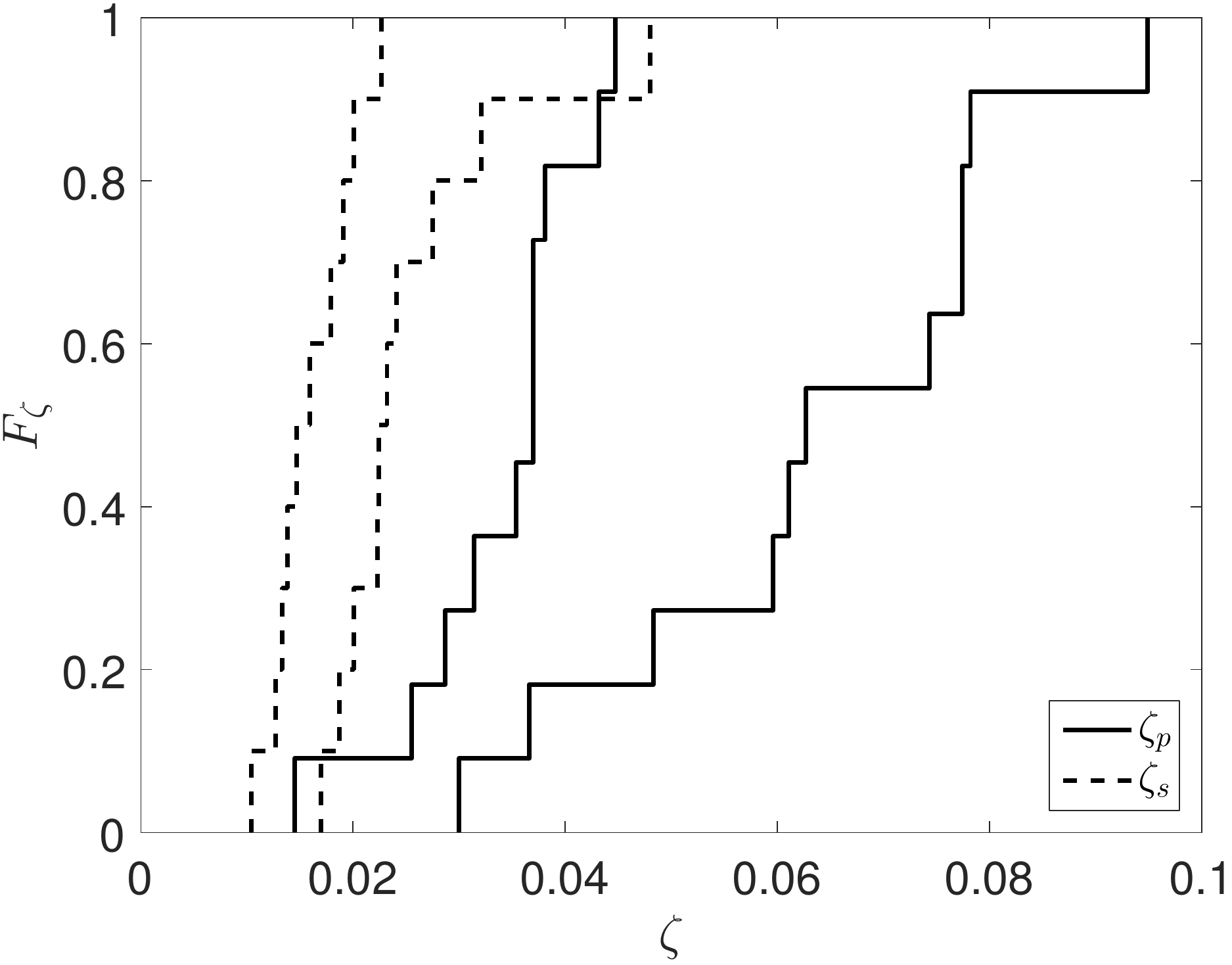}
\caption{\label{fig:2dof:experts} 2-dof damped oscillator -- aggregated p-boxes of the damping ratio (10 expert opinions of equal credibility)}
\end{figure}

Consider the two-degree-of-freedom (2-dof) damped oscillator subjected to white noise excitation $S(t)$, sketched in Figure~\ref{fig:damp:sketch}. The subscripts $p$ and $s$ refer to the primary and secondary mass, respectively. The QoI is the force acting on the secondary spring. According to \cite{DK1990,Dubourg2013}, the peak force $P_s$ in the secondary spring can be computed by:
\begin{equation}
P_s = 3\, k_s \sqrt{\mathbb{E}_S\bra{x_s^2}}, \qquad \text{where} \qquad \mathbb{E}_S\bra{x_s^2} = \pi \frac{S_0}{4\zeta_s\omega_s^3}\, \frac{\zeta_a\zeta_s}{\zeta_p\zeta_s\prt{4\zeta_a^2+\xi^2}+\gamma\zeta_a^2} \, \frac{\prt{\zeta_p\omega_p^3+\zeta_s\omega_s^3}\omega_p}{4\zeta_a\omega_a^4},
\end{equation}
where $\omega_p = \sqrt{k_p/m_p}$ and $\omega_s = \sqrt{k_s/m_s}$ are the natural frequencies, $\gamma = m_s/m_p$, $\omega_a = \prt{\omega_p+\omega_s}/2$, $\zeta_a = \prt{\zeta_p+\zeta_s}/2$, $\xi = \prt{\omega_p-\omega_s}/\omega_a$, $m$ is the mass, $k_p$ and $k_s$ are the spring stiffnesses, and $\zeta$ is the damping ratio. The set of input parameters $\vx = \acc{m_p, m_s, k_p, k_s, \zeta_p, \zeta_s, S_0}$ are considered statistically independent. The parameters $\acc{m_p,m_s,k_p,k_s,S_0}$ are well-known (\ie negligible epistemic uncertainty) and defined as independent lognormal variables, with properties summarized in Table~\ref{tab:damp:init}. Each lognormal variable is characterized by a mean value and a coefficient of variation (CoV). Assume that the properties of the damping ratios $\acc{\zeta_p,\zeta_s}$ are investigated through a survey among experts because of the highly uncertain nature of damping in dynamic systems. Analogously to Case \#1, each expert provides an interval of values for the two damping ratios. Eleven and ten intervals have been collected for $\zeta_p$ and $\zeta_s$, respectively. The aggregated p-boxes are shown in Figure~\ref{fig:2dof:experts} assuming equal credibility among the experts.

\begin{table}[ht!]
\centering
\caption{\label{tab:damp:init} 2-dof damped oscillator -- probabilistic input variables}
\begin{tabular}{llll}
\hline
Variable & Distribution & Mean & CoV \\
\hline
$m_p$ & Lognormal & $1.50$ & $10\%$ \\
$m_s$ & Lognormal & $0.01$ & $10\%$ \\
$k_p$ & Lognormal & $1.00$ & $20\%$\\
$k_s$ & Lognormal & $0.05$ & $20\%$ \\
\hline
$S_0$ & Lognormal & $100$ & $10\%$  \\
\hline
\end{tabular}
\end{table}

\subsubsection{Analysis}
The settings for the two-level meta-modelling approach are kept the same as in the Rosenbrock function example in Section~\ref{sec:appl:2d}. Sparse PCE is trained with a candidate basis of maximum total polynomial degree equal to $20$ and hyperbolic truncation sets with $q=0.75$. The auxiliary variables $\widetilde{X}_i$ for $\zeta_p$ and $\zeta_s$ are defined as in Eq.~(\ref{eq:auxU}). In order to analyse the influence of the sample size, $N_1$ and $N_2$ are varied in $\acc{100, 200, 300, \rightarrow\infty}$ and $\acc{100,200,300,1000}$, respectively. Note that $N_1\rightarrow\infty$ denotes a case where the first-level meta-model is not applied; instead, the true model $\cm$ is used. 50 independent analyses with different experimental designs are conducted to assess the statistical significance of the QoI.

\subsubsection{Results}
The results for the relative generalization error are summarized in Table~\ref{tab:2dof:free:results}. As expected, increasing $N_1$ reduces the relative generalization error of the first-level meta-model (see $\widehat{err}_{\text{gen}}\bra{\widetilde{P}_s}$). However, the accuracy of the second-level meta-model is lower than the first-level meta-model, and does not depend significantly on the accuracy of the first-level meta-model. Thus, the error is dominated by the second-level meta-modelling operation. The explanation lies in the p-boxes of the damping coefficients $\zeta_p$ and $\zeta_s$, which have stairs-shaped boundary CDF curves. Analogously to Case \#1 in the Rosenbrock function (Section~\ref{sec:appl:2d}), this shape reduces the accuracy of the meta-model on the second level.  

\begin{table}[!ht]
\centering
\caption{2-dof damped oscillator -- resulting relative generalization error $\widehat{err}_{\text{gen}}$ based on a Monte Carlo simulation with $n=10^5$ samples -- mean value of 50 repetitions \label{tab:2dof:free:results}}
\begin{tabular}{lllll} 
\hline
 $N_1$ & $\widehat{err}_{\text{gen}}\bra{\widetilde{P}_s}$ & $N_2$ & $\widehat{err}_{\text{gen}}\bra{\underline{P}_s}$  &  $\widehat{err}_{\text{gen}}\bra{\overline{P}_s}$ \\
\hline
$100$ & $1.59\cdot 10^{-2}$ & $100$ & $5.85\cdot 10^{-2}$ & $9.46\cdot 10^{-2}$\\
& & $200$ & $4.59\cdot 10^{-2}$ & $7.32\cdot 10^{-2}$\\
& &  $300$ & $4.18\cdot 10^{-2}$& $6.76\cdot 10^{-2}$ \\
& & $1000$ & $3.19\cdot 10^{-2}$& $5.17\cdot 10^{-2}$ \\
\hline
$200$ & $5.24\cdot 10^{-3}$ & $100$ & $5.20\cdot 10^{-2}$ & $7.99\cdot10^{-2}$\\
&  & $200$ & $3.41\cdot 10^{-2}$& $6.08\cdot 10^{-2}$\\
&  & $300$ & $2.97\cdot 10^{-2}$ & $5.21\cdot 10^{-2}$\\
&  &$1000$ & $1.87\cdot 10^{-2}$& $3.57\cdot 10^{-2}$\\
\hline
$300$ & $2.85\cdot 10^{-3}$ & $100$ & $5.12\cdot 10^{-2}$ & $8.09\cdot 10^{-2}$ \\
& &  $200$ & $3.13\cdot10^{-2}$ & $5.93\cdot10^{-2}$ \\
& &  $300$ & $2.75\cdot10^{-2}$& $4.99\cdot10^{-2}$\\
& & $1000$ & $1.71\cdot 10^{-2}$& $3.30\cdot 10^{-2}$\\
\hline
$\rightarrow\infty$ & $0$ & $100$ & $5.44\cdot 10^{-2}$ & $8.49\cdot 10^{-2}$ \\
& &  $200$ & $3.07\cdot10^{-2}$& $6.11\cdot 10^{-2}$\\
& &  $300$ & $2.71\cdot 10^{-2}$ & $4.96\cdot 10^{-2}$ \\
& & $1000$ & $1.52\cdot 10^{-2}$& $3.07\cdot 10^{-2}$\\
\hline
\end{tabular}
\end{table}

\begin{figure}[!ht]
\centering
	\includegraphics[width=0.45\linewidth]{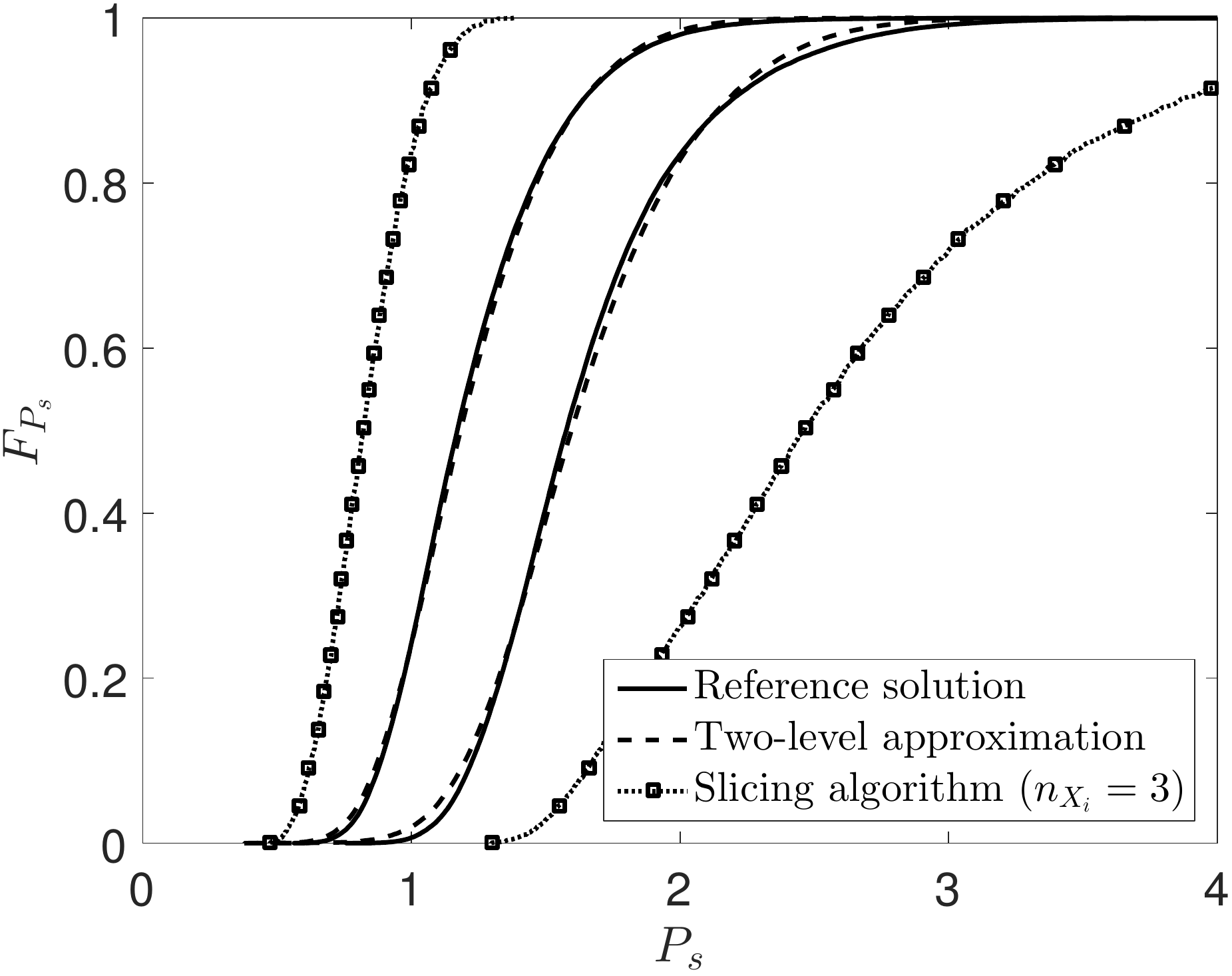}
\caption{2-dof damped oscillator -- QoI $P_s$: reference solution ($n=10^5$) versus two-level approximation approach ($N_1 = 300$, $N_2 = 100$, $\widehat{err}_{\text{gen}}\bra{\widetilde{P}_s} = 2.57\cdot 10^{-3}$, $\widehat{err}_{\text{gen}}\bra{\overline{P}_s} = 7.50\cdot 10^{-2}$, $\widehat{err}_{\text{gen}}\bra{\underline{P}_s} = 3.89\cdot 10^{-2}$) and slicing algorithm with $n_{X_i}=3$ \label{fig:2dof}}
\end{figure}

Although the two-level meta-models are fairly accurate, as discussed previously, the bounds of the resulting p-box is computed accurately, as seen in Figure~\ref{fig:2dof}. A reference solution based on $n=10^5$ Monte Carlo samples is compared to the two-level meta-modelling approach with $N_1 = 300$ and $N_2=100$. The boundary curves ($\lF_{P_s}$ and $\uF_{P_s}$) obtained from the proposed approach are almost superimposed with the reference ones. 

\subsubsection{Comparison to slicing algorithm}
\begin{figure}[ht!]
\centering
\subfigure{
\includegraphics[width=0.3\linewidth]{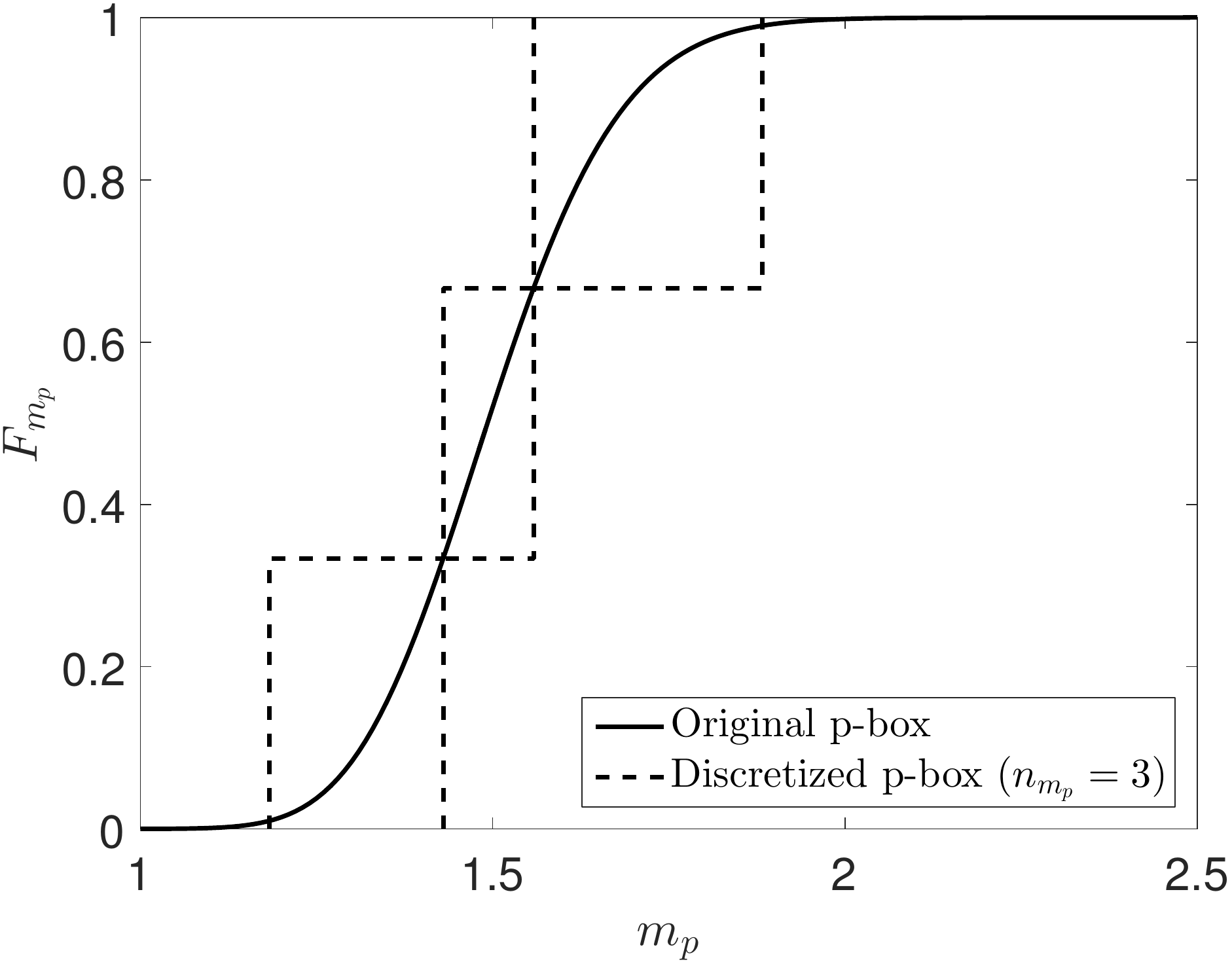}
}
\subfigure{
\includegraphics[width=0.3\linewidth]{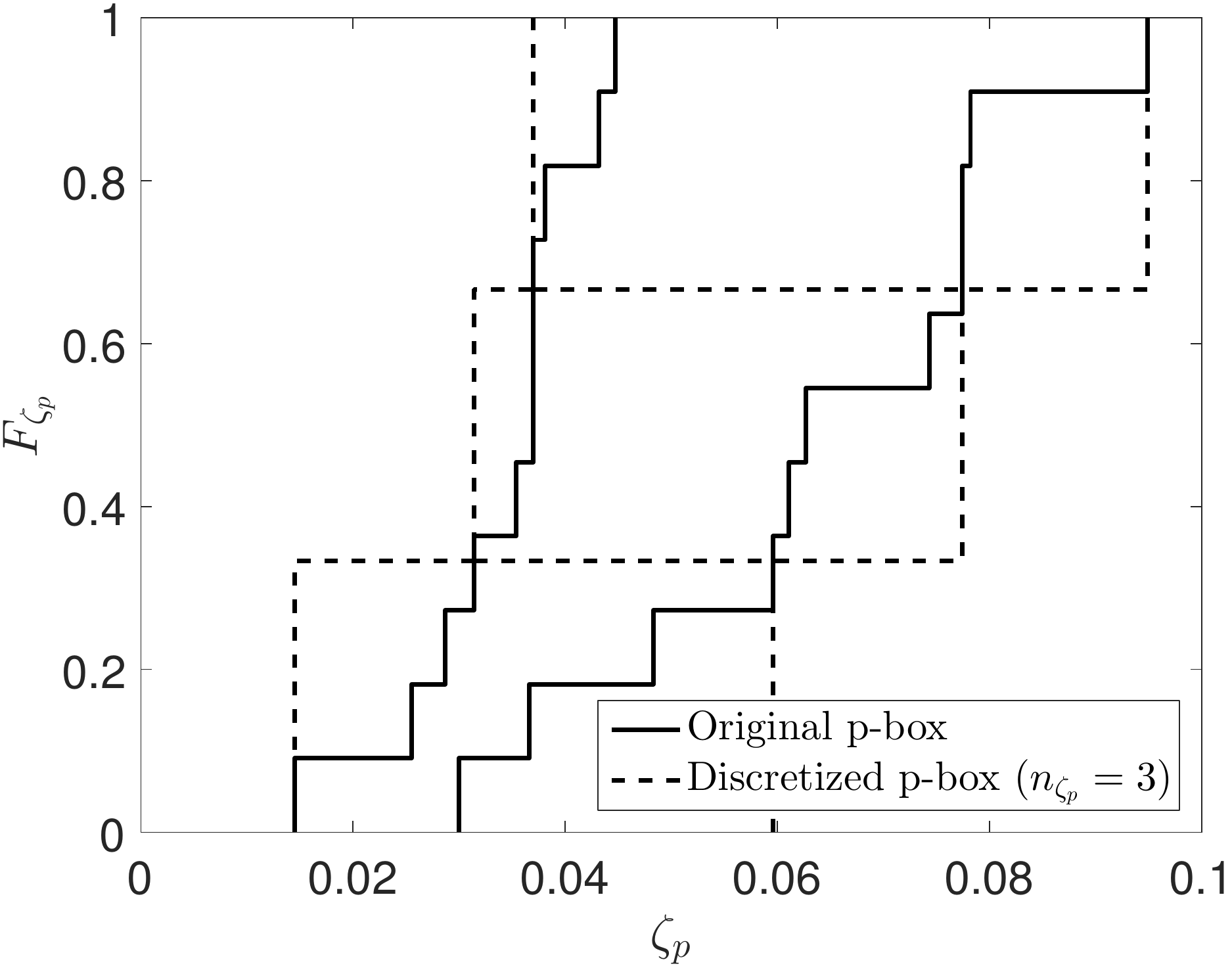}
}
\caption{\label{fig:2dof:slice} 2-dof damped oscillator -- discretization of $m_p$ and $\zeta_p$ with $n_{X_i}=3$}
\end{figure}

In order to illustrate the curse of dimensionality, the slicing algorithm is applied on the second level of the two-level approach replacing the calibration of the meta-models $\lcm^{(\text{P})}$ and $\ucm^{(\text{P})}$. Note that in order to apply the slicing algorithm consistently, each probabilistic input variable is interpreted as p-box too. These continuous random variables are bounded within their 1\% and 99\% quantiles in order to obtain finite intervals for the interval analysis. The number of intervals is chosen as $n_{X_1} = 3$ in order to obtain a similar number of optimization operations as the proposed two-level approach, \ie $|\cK|=3^7=2187$. Figure~\ref{fig:2dof:slice} illustrates the discretization for $m_p$ and $\zeta_p$. The boundary curves of the discretized p-boxes are wider than the original input p-boxes. This effect is pronounced probabilistic input variables, as seen in the left part of Figure~\ref{fig:2dof:slice}. 

The discretization algorithm leads to conservative approximations of the true input p-boxes. The propagation of the discretized intervals results in a p-box that is much wider (\ie more conservative) compared to the proposed two-level approach, as indicated in Figure~\ref{fig:2dof}, despite the fact that more model evaluations were computed. The conservative estimate of the response p-box originate mainly in the coarse approximation of the five probabilistic input variables.  

\subsection{Two-dimensional truss structure} \label{sec:appl:truss}

\subsubsection{Problem definition}
\begin{figure}[ht!]
\centering
\includegraphics[width=0.6\linewidth]{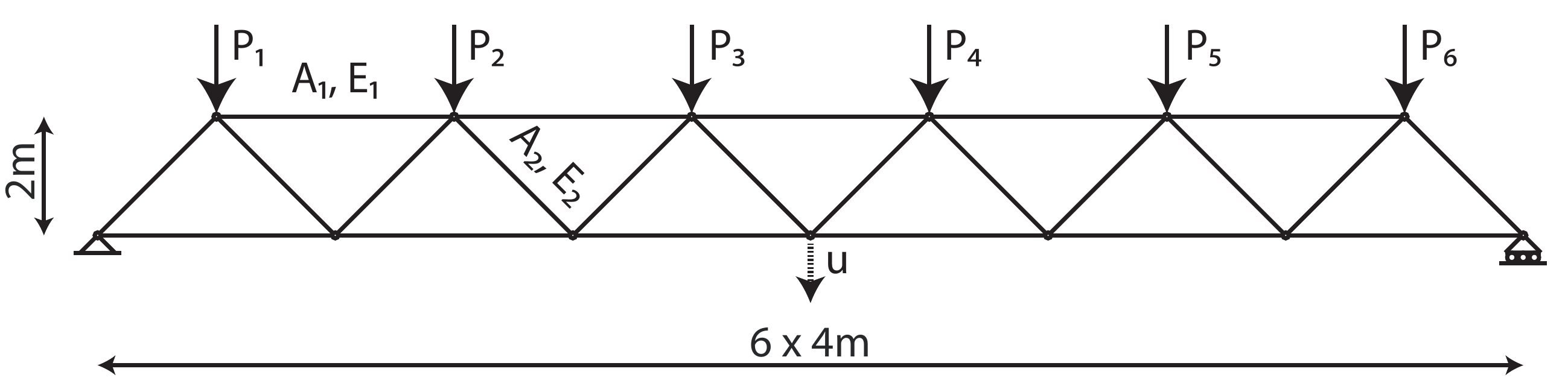}
\caption{\label{fig:truss:sketch} Two-dimensional truss -- geometry, loading and QoI}
\end{figure}

The third application example is a two-dimensional truss structure, which has been discussed previously in the context of reliability analysis \citep{Lee:Kwak:2006,BlatmanPEM2010,SchobiASCE2015}. Figure~\ref{fig:truss:sketch} shows the truss which is composed of 13 nodes and 23 bars. The geometry of the nodes is given deterministically, whereas the loading and material properties are given by p-boxes. The upper chord of the truss is subjected to loads $P_i$, $i=1,\ldots,6$. The horizontal bars are described by the cross-sectional area $A_1$ and the Young's modulus $E_1$ whereas the diagonal bars are described by $A_2$ and $E_2$, respectively. The 10-dimensional input vector then reads:
\begin{equation}
\vX = \bra{ A_1,\,A_2,\,E_1,\,E_2,\,P_1,\,P_2,\,P_3,\,P_4,\,P_5,\,P_6}.
\end{equation}
Analogously to Case \#2(b), the input variables are modelled by p-boxes defined by the envelope of a set of curves. The distribution function and its interval-valued parameters are summarized in Table~\ref{tab:truss:input}. The variables are assumed to be statistically independent.  

\begin{table}[ht!]
\centering
\caption{\label{tab:truss:input} Two-dimensional truss -- input p-boxes}
\begin{tabular}{lllll}
\hline
Variable & & Distribution & Mean & CoV \\
\hline
$A_1$ & [m$^2$] & Lognormal & $[1.9,\,2.1]\cdot 10^{-3}$ & $[8,\,12]\%$\\
$A_2$ & [m$^2$] & Lognormal & $[0.9,\,1.1]\cdot 10^{-3}$ & $[8,\,12]\%$\\
$E_1$, $E_2$ & [Pa] & Lognormal & $[2.0,\,2.2]\cdot 10^{11}$ & $[9,\,11]\%$\\
$P_1,\ldots,P_6$ & [N] & Gumbel & $[4.0,\,6.0]\cdot 10^4$& $[10,\,15]\%$ \\
\hline 
\end{tabular}
\end{table}

The QoI is the deflection $u$ at mid-span as a function of the loading and the material parameters. The arrow in Figure~\ref{fig:truss:sketch} indicates positive values for $u$.

\subsubsection{Analysis}
For the truss structure, the auxiliary variable $\widetilde{X}_i$ is defined as follows. The mean value $\mu_{\widetilde{X}_i}$ is the mid-range of the mean value of $X_i$, and the coefficient of variation $CoV_{\widetilde{X}_i}$ is set to the maximum value of the the coefficient of variation of $X_i$. The auxiliary variables are used on both levels of meta-models to ensure a good convergence behaviour. 

Similar to the previous examples, the number of samples in varied in both levels of meta-models, \ie $N_1=\acc{100,300, \rightarrow\infty}$ and $N_2=\acc{100,300}$. The sparse PCE meta-models are built with Hermite polynomials with a candidate basis of maximal total degree of 20 and hyperbolic truncation scheme with $q=0.75$. To ensure statistical significance of the results, 50 independent runs of the analysis are performed with different LHS experimental designs.

\subsubsection{Results}

A summary of the relative generalization error is given in Table~\ref{tab:truss:free:results}. As expected, increasing the number of samples in the experimental design decreases the error of the meta-model on both meta-modelling levels. In particular, the influence of a larger experimental design on the first-level meta-model is visible. For the second-level meta-model, the absolute value of the relative generalization error is larger than for the first level due to the larger complexity of the analysis. In fact, the accuracy of $\underline{u}$ and $\overline{u}$ depends on the quality of (i) the first-level meta-model, (ii) the optimization algorithm, and (iii) the second-level meta-model.

\begin{table}[!ht]
\centering
\caption{Two-dimensional truss -- resulting relative generalization error based on a Monte Carlo simulation with $n=10^5$ samples -- mean value of 50 repetitions \label{tab:truss:free:results}}
\begin{tabular}{lllll} 
\hline
 $N_{1}$  & $\widehat{err}_{\text{gen}}\bra{\widetilde{u}}$ & $N_{2}$ & $\widehat{err}_{\text{gen}}\bra{\underline{u}}$ & $\widehat{err}_{\text{gen}}\bra{\overline{u}}$ \\
\hline
$100$ & $1.96\cdot 10^{-3}$ & $100$ &  $8.30\cdot 10^{-2}$ & $4.85\cdot 10^{-2}$ \\
 && $300$ &  $2.39\cdot 10^{-2}$ & $2.50\cdot 10^{-2}$ \\
\hline
$300$ & $1.44\cdot 10^{-4}$& $100$ & $9.10\cdot 10^{-2}$& $4.66\cdot 10^{-2}$ \\
 && $300$ & $2.09\cdot 10^{-2}$ & $2.01\cdot 10^{-2}$ \\
\hline
$\rightarrow\infty$ &$0$& $100$ & $1.24\cdot 10^{-2}$ & $2.18\cdot 10^{-2}$ \\
 && $300$ & $7.05\cdot 10^{-3}$ & $1.65\cdot 10^{-2}$ \\
\hline
\end{tabular}
\end{table}

\begin{figure}[!ht]
\centering
\includegraphics[width=0.45\linewidth]{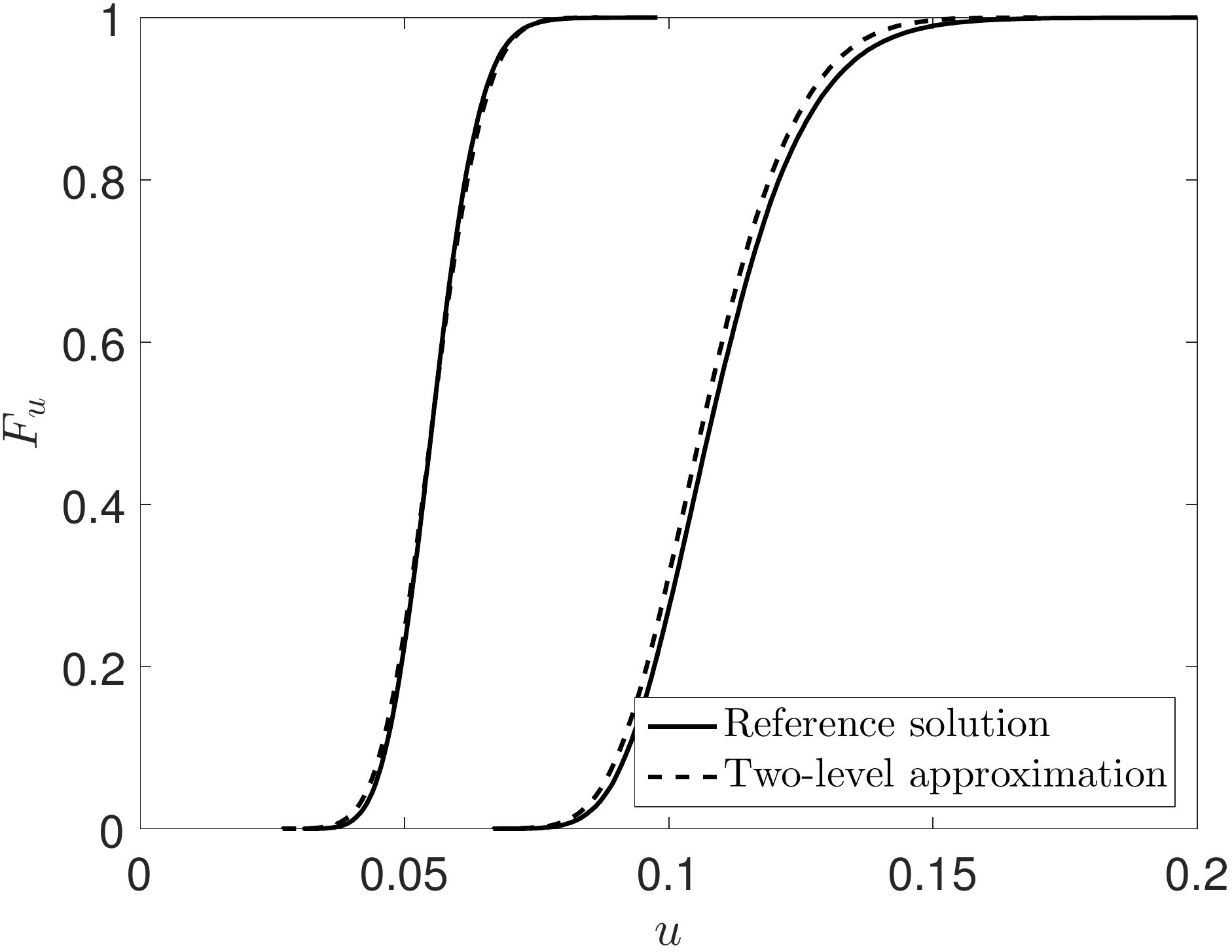}
\caption{Two-dimensional truss -- p-box of the deflection $u$: reference solution ($n=10^5$) versus two-level meta-modelling approach ($N_1 = 300$, $N_2 = 300$) ($\widehat{err}_{\text{gen}}\bra{\widetilde{u}} = 9.82\cdot 10^{-5}$, $\widehat{err}_{\text{gen}}\bra{\overline{u}} = 3.70\cdot 10^{-2}$, and $\widehat{err}_{\text{gen}}\bra{\underline{u}} = 2.64\cdot 10^{-2}$) \label{fig:truss}}
\end{figure}

Another aspect of the modelling is the shape of the response p-box, which is shown in Figure~\ref{fig:truss}. The reference solution is obtained by a Monte Carlo simulation ($n=10^5$) and by taking advantage of the monotonicity of the truss model. A two-level approximation with  $N_1=300$ is able to reproduce the boundary curves of the response p-box accurately: the two curves for $\overline{u}$ coincide, whereas the two curves for $\underline{u}$ are remarkably close to each other, too.

\section{Conclusions} \label{sec:conc}

This paper deals with the propagation of uncertainty in the input parameters of a deterministic, black-box computational model. Traditionally, the uncertainty of the input variables is described using probability theory, \ie by probability distributions. In engineering practice however, data may be too limited to allow for an accurate, purely probabilistic modelling. This can be accounted for by modelling the input parameters using imprecise probability theory, which accounts for both aleatory (natural variability) and epistemic uncertainty (lack of knowledge). The use of probability-boxes (p-boxes) is one way to capture this mixed uncertainty by providing lower and upper bounds for the cumulative distribution function of variables. 

In the context of p-boxes modelling, uncertainty propagation analyses are not straightforward. Using a simple problem conversion, the problem is recast as two independent uncertainty propagation problems based on standard probabilistic modelling of the inputs. This problem conversion allows for traditional sampling-based methods, such as Monte Carlo simulation. However, due to the repeated evaluation of the computational model in sampling-based methods, the computational costs may become intractable. 

In this paper, a two-level meta-modelling approach is proposed to reduce the overall computational cost considerably. Sparse non-intrusive Polynomial Chaos Expansions (PCE) are used to surrogate the computational model. Sparse PCE allows for an efficient (accurate and inexpensive) modelling of the bounds of the response p-box.

The capabilities of the two-level approach are illustrated by a benchmark analytical function and two realistic engineering problems. In all examples, the proposed approach is capable of estimating the bounds of the response p-box accurately with only a small number of runs of the exact computational model. This is of major significance in practice where computational resources are typically limited and at the same time expensive-to-evaluate models are analysed.

\section*{References}

\bibliographystyle{chicago}
\bibliography{biblioIUQ}

\end{document}